\documentclass[
               letterpaper,%
               twoside,%
               10pt%
               ]{article}

\usepackage[hmargin={1.0in,1.0in},
            vmargin={1.0in,1.0in},%
            headheight=14pt,
            ]{geometry}

\usepackage{amsmath}
\numberwithin{equation}{section}                        
\usepackage{amsthm}
\usepackage{amssymb}                                    
\usepackage{bbm}                                        
\makeatletter
\let\@upn\@iden
\makeatother
\newtheoremstyle{ToddTheorem}                           
  {3pt}
  {3pt}
  {\slshape}
  {}
  {\bfseries\upshape}
  {.}
  {.5em}
  {}

\newtheoremstyle{ToddDefinition}                        
  {3pt}
  {3pt}
  {\upshape}
  {}
  {\bfseries\slshape}
  {.}
  {.5em}
  {}

\makeatletter
\newcommand*{\c@theoremassumption}{\c@theorem}
\newcommand*{\p@theoremassumption}{\p@theorem}

\makeatother

\makeatletter
\newcommand*{\c@theoremconjecture}{\c@theorem}
\newcommand*{\p@theoremconjecture}{\p@theorem}

\makeatother

\makeatletter
\newcommand*{\c@theoremcorollary}{\c@theorem}
\newcommand*{\p@theoremcorollary}{\p@theorem}

\makeatother

\makeatletter
\newcommand*{\c@theoremdefinition}{\c@theorem}
\newcommand*{\p@theoremdefinition}{\p@theorem}

\makeatother

\makeatletter
\newcommand*{\c@theoremexample}{\c@theorem}
\newcommand*{\p@theoremexample}{\p@theorem}

\makeatother

\makeatletter
\newcommand*{\c@theoremfigure}{\c@theorem}
\newcommand*{\p@theoremfigure}{\p@theorem}

\makeatother

\makeatletter
\newcommand*{\c@theoremhypothesis}{\c@theorem}
\newcommand*{\p@theoremhypothesis}{\p@theorem}

\makeatother

\makeatletter
\newcommand*{\c@theoremlemma}{\c@theorem}
\newcommand*{\p@theoremlemma}{\p@theorem}

\makeatother

\makeatletter
\newcommand*{\c@theoremproposition}{\c@theorem}
\newcommand*{\p@theoremproposition}{\p@theorem}

\makeatother

\makeatletter
\newcommand*{\c@theoremremark}{\c@theorem}
\newcommand*{\p@theoremremark}{\p@theorem}

\makeatother

\makeatletter
\newcommand*{\c@theoremnotation}{\c@theorem}
\newcommand*{\p@theoremnotation}{\p@theorem}

\makeatother

\newtheoremstyle{SpringerTheorem}                           
  {3pt}
  {3pt}
  {\itshape}
  {}
  {\bfseries}
  {.}
  {.5em}
  {}
\newtheoremstyle{SpringerDefinition}                        
  {3pt}
  {3pt}
  {\rmfamily}
  {}
  {\bfseries}
  {.}
  {.5em}
  {}
\newtheoremstyle{SpringerExample}                        
  {3pt}
  {3pt}
  {\rmfamily}
  {}
  {\itshape}
  {.}
  {.5em}
  {}

\theoremstyle{SpringerTheorem} %

\newtheorem{lemma}[theoremlemma]{Lemma}

\theoremstyle{SpringerDefinition} %

\theoremstyle{SpringerExample} %

\newtheorem{remark}[theoremremark]{Remark}

\newcommand{\TheAuthor}{}
\newcommand{\Author}[1]%
        {\renewcommand{\TheAuthor}{#1}}                 
\newcommand{\TheRunningTitle}{}
\newcommand{\RunningTitle}[1]%
        {\renewcommand{\TheRunningTitle}{#1}}           

\usepackage{fancyhdr}                                   
\renewcommand{\footrulewidth}{0.4pt}

\usepackage{titling}
\pretitle{\begin{center}\LARGE\sffamily\slshape}        
\posttitle{\par\end{center}\vskip 0.5em}
\predate{\begin{center}\small}                          
\postdate{\par\end{center}\vskip 0.5em}
\thanksmarkseries{fnsymbol}                             

\usepackage[runin]{abstract}
\abslabeldelim{.\,}

\usepackage[titles]{tocloft}                            

\usepackage{titlesec}
\titleformat{\section}%
   {\Large\scshape\filcenter}{\thesection.}{1em}{}
\titleformat{\subsection}%
   {\large\sffamily\slshape\filcenter}{\thesubsection.}{1em}{}
\titleformat{\subsubsection}%
   {\normalsize\sffamily\filcenter}{\thesubsubsection.}{1em}{}
\titlespacing*{\section}%
   {0pt}{4.5ex plus 1ex minus .2ex}{3.3ex plus .2ex}
\titlespacing*{\subsection}%
   {0pt}{4.25ex plus 1ex minus .2ex}{3.05ex plus .2ex}
\titlespacing*{\subsubsection}%
   {0pt}{4.25ex plus 1ex minus .2ex}{3.05ex plus .2ex}

\usepackage[comma,numbers,square,sort&compress]{natbib} 

\usepackage{bm}                                         

\usepackage[
            ]{graphicx}                                 

\usepackage[
            small,%
            bf]{caption}                     
\setcaptionmargin{0.5in}

\newenvironment{acknowledgment}%
  {\begin{trivlist}\item[]\textbf{Acknowledgments.}}{\end{trivlist}}

\makeatletter
\providecommand*{\toclevel@assumption}{0}
\providecommand*{\toclevel@conjecture}{0}
\providecommand*{\toclevel@corollary}{0}
\providecommand*{\toclevel@definition}{0}
\providecommand*{\toclevel@example}{0}
\providecommand*{\toclevel@figure}{0}
\providecommand*{\toclevel@hypothesis}{0}
\providecommand*{\toclevel@lemma}{0}
\providecommand*{\toclevel@proof}{0}
\providecommand*{\toclevel@proposition}{0}
\providecommand*{\toclevel@remark}{0}
\providecommand*{\toclevel@theorem}{0}
\makeatother

\newcommand{\eref}[1]{\hyperref[{#1}]{(\ref*{#1})}}

\newcommand{\C}{\mathbb{C}}

\newcommand{\E}{\mathbb{E}}

\newcommand{\R}{\mathbb{R}}

\def\arg{\mathop\mathrm{arg}\nolimits}

\def\det{\mathop\mathrm{det}\nolimits}
\def\diag{\mathop\mathrm{diag}\nolimits}
\def\dim{\mathop\mathrm{dim}\nolimits}

\def\ker{\mathop\mathrm{ker}\nolimits}

\def\res{\mathop\mathrm{Res}\limits}

\def\sgn{\mathop\mathrm{sign}\nolimits}
\def\Span{\mathop\mathrm{span}\nolimits}

\def\Re{\mathop\mathrm{Re}\nolimits}
\def\Im{\mathop\mathrm{Im}\nolimits}

\def\coloneqq{\mathrel{\mathop:}=}

\newcommand{\rma}{\mathrm{a}}

\newcommand{\rmc}{\mathrm{c}}
\newcommand{\rmd}{\mathrm{d}}
\newcommand{\rme}{\mathrm{e}}

\newcommand{\rmg}{\mathrm{g}}

\newcommand{\rmi}{\mathrm{i}}

\newcommand{\rmm}{\mathrm{m}}
\newcommand{\rmn}{\mathrm{n}}

\newcommand{\rmp}{\mathrm{p}}
\newcommand{\rmr}{\mathrm{r}}

\newcommand{\rmT}{\mathrm{T}}

\newcommand{\Ham}{\mathrm{Ham}}
\newcommand{\calH}{\mathcal{H}}
\newcommand{\calI}{\mathcal{I}}
\newcommand{\calJ}{\mathcal{J}}

\newcommand{\calL}{\mathcal{L}}

\newcommand{\calO}{\mathcal{O}}

\newcommand{\calR}{\mathcal{R}}
\newcommand{\calS}{\mathcal{S}}
\newcommand{\calT}{\mathcal{T}}

\newcommand{\vD}{\bm{\mathit{D}}}

\newcommand{\vK}{\bm{\mathit{K}}}

\newcommand{\vc}{\bm{\mathit{c}}}

\newcommand{\vu}{\bm{\mathit{u}}}

\newcommand{\vn}{\bm{\mathit{0}}}

\Author{T. Kapitula, P. Kevrekidis, and D. Yan}
\RunningTitle{Krein Matrix: Theory and Applications}

\usepackage[letterpaper,%
            bookmarks=false,%
            bookmarksopen=false,%
            bookmarksnumbered=false,%
            hypertexnames=false,%
            colorlinks=true,%
            linkcolor=blue,%
            citecolor=blue,%
            urlcolor=blue]{hyperref}                    

\usepackage{kpfonts}

\begin{document}

\title{The Krein Matrix: General Theory and Concrete Applications in
Atomic Bose-Einstein Condensates }
\author{%
        Todd Kapitula %
        \thanks{E-mail: \href{mailto:tmk5@calvin.edu}{tmk5@calvin.edu}} \\
        Department of Mathematics and Statistics \\
        Calvin College \\
        Grand Rapids, MI 49546 %
\and
        Panayotis G. Kevrekidis%
         \thanks{E-mail: \href{mailto:kevrekid@math.umass.edu}{kevrekid@math.umass.edu}} \\
         Department of Mathematics and Statistics\\ %
         University of Massachusetts\\ %
         Amherst, MA 01003-4515 %
\and
        Dong Yan%
         \thanks{E-mail: \href{mailto:yan@math.umass.edu}{yan@math.umass.edu}} \\
         Department of Mathematics and Statistics\\ %
         University of Massachusetts\\ %
         Amherst, MA 01003-4515 %
}


\begin{titlingpage}
\usethanksrule
\setcounter{page}{0}                                    
\maketitle
\begin{abstract}
When finding the nonzero eigenvalues for Hamiltonian eigenvalue problems it
is especially important to locate not only the unstable eigenvalues (i.e.,
those with positive real part), but also those which are purely imaginary but
have negative Krein signature. These latter eigenvalues have the property
that they can become unstable upon collision with other purely imaginary
eigenvalues, i.e., they are a necessary building block in the mechanism
leading to the so-called Hamiltonian-Hopf bifurcation. In this paper we
review a general theory for constructing a meromorphic matrix-valued
function, the so-called Krein matrix, which has the property of not only
locating the unstable eigenvalues, but also those with negative Krein
signature. These eigenvalues are realized as zeros of the determinant. The
resulting finite dimensional problem obtained by setting the determinant of
the Krein matrix to zero presents a valuable simplification.
In this paper the usefulness
of the technique is illustrated through prototypical examples of spectral
analysis of states that have arisen in recent experimental and theoretical
studies of atomic Bose-Einstein condensates. In particular, we consider
one-dimensional settings (the cigar trap) possessing real-valued
multi-dark-soliton solutions, and two-dimensional settings (the pancake
trap) admitting complex multi-vortex stationary waveforms.
\end{abstract}

\cancelthanksrule
\renewcommand{\footrulewidth}{0.0pt}                    

\pdfbookmark[1]{\contentsname}{toc}                     
\tableofcontents                                        
\end{titlingpage}

\section{Introduction}\label{s:intro}

Hamiltonian eigenvalue problems have a time-honored history, as they arise in
numerous applications stemming from fluid mechanics, celestial mechanics,
optical and atomic physics among many other disciplines; see for some recent
examples the books~\cite{kev,peli,yang}. Especially in higher dimensional
settings these problems can rapidly become fairly computationally
intractable, at least as concerns providing the full diagonalization of the
relevant matrix (e.g., when it stems from the linearization around
two-dimensional vortex structures or three-dimensional
vortex-rings~\cite{komineas}). It is therefore highly desirable to be able to
reduce the dimensionality of the calculation
by providing a technique that can capture the main features of the
linearization spectrum through suitable reductions to a finite dimensional
eigenvalue problem. It is the aim of the present work to provide a general
overview, as well as a systematic set of case examples of such a method. The
approach that will be developed will be based on the so-called Krein
matrix~\cite{kapitula:tks10}. The Krein matrix is a meromorphic matrix-valued
function constructed via a Lyapunov-Schmidt reduction, and consequently
recasts the infinite-dimensional eigenvalues problem as a finite-dimensional
one. The construction is such that the eigenvalues are realized as points for
which the Krein matrix is singular. The computation and visualization of the
determinant of this matrix can serve as a tool to identify the eigenvalues of
the original problem.

When determining the spectrum for the Hamiltonian eigenvalue problem, there
are two spectral sets to consider: those with positive real part, and those
with negative Krein index (signature). The latter eigenvalues are purely
imaginary; however, upon collision with eigenvalues of positive Krein
signature it will generically be the case that a Hamiltonian-Hopf bifurcation
will occur, which leads to an oscillatory instability. While the eigenvalues
with positive real part are easy to visually identify, an additional
calculation is necessary in order to identify the signature of a purely
imaginary eigenvalue. As we will see later in this paper, the Krein matrix is
constructed in such a manner that the eigenvalues with negative signature can
be identified \textit{graphically}. Consequently, all of the (potentially)
unstable eigenvalues can be identified visually.

A related question is: how many (potentially) unstable eigenvalues are there
to locate? For a given problem it may be theoretically possible to establish
an upper bound on the real part of all eigenvalues; however, in many problems
of interest there is no upper bound on the imaginary part of the eigenvalue.
The Hamiltonian-Krein index, which will be discussed in detail later in this
paper, counts the number of eigenvalues with positive real part, as well as
the number of purely imaginary eigenvalues with negative Krein signature. In
the problems presented in this paper this index will be finite, and it will
be related to the number of negative directions of the constrained second
variation of the energy (the underlying wave is realized as a critical point
of the constrained energy).  Thus, while no bound is present on the imaginary
part of all of the eigenvalues, since there are only a finite number of
(potentially) unstable eigenvalues, there will be an upper bound for this
set. A theoretical determination of this bound is probably not possible;
however, it can be determined numerically through the Krein matrix.

The computation of the Krein matrix in this paper will be numerical for each
case study. Unfortunately, at this point in time we do not know of any
examples for which the Krein matrix can be explicitly computed. Our hope is
that for special problems, e.g., the nonlinear Schr\"odinger equation with
the 1-soliton potential, such a calculation may be possible. This will be the
topic of future research. We note that there is reason to be optimistic that
this will be a fruitful direction for research, in that it is possible in
some special cases to explicitly construct the Evans function (another
eigenvalue counting tool), e.g., see \citep[Chapters~9.3 and
10.4]{kapitula:sad13} and the references therein.

Our presentation will be structured as follows. In \autoref{s:hamspectra}, we
first provide an overview of the Hamiltonian-Krein index theory for
Hamiltonian eigenvalue problems. Afterwards, we focus on the construction of
the Krein matrix and provide a summary of its properties. In \autoref{s:3} we
consider specific examples stemming from the application of the Krein matrix
analysis to vortex dynamical states that are of intense recent interest in
the field of atomic physics. In particular, we consider single (unit charge)
vortex states that are presently fairly routine to
produce/observe~\cite{emergent}, but also examine the case of the recently
studied vortex dipoles. For the sake of completeness, we also examine in
\autoref{s:4} one-dimensional analogs of such vortex states; namely,
multi-dark soliton structures that have been studied theoretically (see
e.g.~\cite{coles,rev} and references therein) and also have been observed
experimentally in~\cite{heid1,heid2}. Finally, in \autoref{s:5} we summarize
our findings and present our conclusions, as well as some potential topics
for future study.

Regarding the states studied in \autoref{s:3}, it is known that the
precessional dynamics of a single vortex is connected with the negative Krein
signature eigenvalues of the corresponding linearization spectrum, see
\citep{dsh1}. The vortex dipoles were produced by dynamical experiments
involving flow past an obstacle~\cite{bpa}, or quenching through the
Bose-Einstein condensation (BEC) transition~\cite{dsh1}. Furthermore, their
dynamical properties and structural robustness were studied in considerable
length in recent works in the physical literature
\citep{kom0,dsh2,dsh3,peli4}. Experimental works on this theme - especially,
\citep{dsh2} - were concerned with issues such as:
\begin{enumerate}
\item the equilibrium configuration, which is also explored herein;
\item the epicyclic motion around this equilibrium, which was analyzed
    through the negative Krein signature eigenvalues of the dipole's
    linearization spectrum.
\end{enumerate}
It should be noted in passing that the study of spectral stability properties
in BEC settings (and particularly for single- and multi-charge vortices) is a
subject rapidly gaining momentum, as evidenced by the recent work of
\citep{kollar:sso12} on the subject.

\begin{acknowledgment}
TK gratefully acknowledges the support of the Jack and Lois Kuipers Applied
Mathematics Endowment, a Calvin Research Fellowship, and the National Science
Foundation under grant DMS-1108783. PGK and DY gratefully acknowledge support
from AFOSR grant FA9550-12-1-0332, from NSF-DMS grant of number 0806762,
and PGK also from the Alexander von Humboldt Foundation and the Binational
Science Foundation, grant number 2010239.
\end{acknowledgment}

\section{Hamiltonian spectral theory}\label{s:hamspectra}

Consider the Hamiltonian eigenvalue problem
\begin{equation}\label{e:11}
\calJ\calL u=\lambda u,
\end{equation}
acting on a Hilbert space $X$ with inner-product $\langle\cdot,\cdot\rangle$.
The operator $\calJ$ is skew-symmetric bounded operator with a bounded
inverse, and the operator $\calL:Y\mapsto X$ is Hermitian with a compact
resolvent. The space $Y\subset X$ is assumed to be dense. Under these
assumptions it is well-known that the spectra of $\calJ\calL$, namely
$\sigma(\calJ\calL)$, is all point spectra, each eigenvalue has finite
multiplicity, and infinity is the only possible accumulation point of the
eigenvalues. If we further assume that each of the operators has zero
imaginary part, i.e., $\Im(\calJ)=\Im(\calL)=0$, then the eigenvalues satisfy
the quartet symmetry that if $\lambda\in\sigma(\calJ\calL)$, then the set
$\{\pm\lambda,\pm\overline{\lambda}\}\subset\sigma(\calJ\calL)$. Furthermore,
the algebraic multiplicities of each of the eigenvalues in the quartet
matches, e.g., $\rmm_\rma(\lambda)=\rmm_\rma(\overline{\lambda})$.

\subsection{The Hamiltonian-Krein index}\label{s:kreinindex}

The Hamiltonian-Krein index theory has a long history (see
\citet{deconinck:ots10,haragus:ots08,kapitula:ace05,kapitula:cev04,pelinovsky:ilf05,chugunova:coe10}
and the references therein). The index theory is used to relate
$\rmn(\calL)$, which is the total number (including multiplicity) of negative
eigenvalues of $\calL$, to the total number of eigenvalues
$\lambda\in\sigma(\calJ\calL)$ with positive real part. In general, for
Hermitian operators $\calH$ we will denote the number of negative eigenvalues
including multiplicity) by $\rmn(\calH)$.

The details for the following discussion can be found in, e.g.,
\cite{deconinck:ots10}, and an abbreviated version is included here for the
sake of completeness. The eigenvalue problem \eref{e:11} arises from
linearizing a particular Hamiltonian system about some type of steady state
solution (solitary wave, spatially periodic wave, etc.). The underlying
system has $N$ symmetries, which means that $\dim[\ker(\calL)]\ge N$, as each
symmetry generates an element of the kernel, and the kernel elements
generated in this fashion are linearly independent. We will henceforth assume
that $\dim[\ker(\calL)]=N$, with $\ker(\calL)=\Span\{\phi_1,\dots,\phi_N\}$.
Now, the generalized kernel is found by solving $\calL u=\calJ^{-1}\phi$,
where $\phi\in\ker(\calL)$. Now, the symmetries generate conserved
quantities, and it turns out to be the case the manner in which these
quantities are generated allows us to solve $\calL u=\calJ^{-1}\phi_j$ for
each $j=1,\dots,N$: denote these solutions as $\psi_j$ (see
\citet{grillakis:sto90,grillakis:sto87}). Thus, $\rmm_\rmg(0)=N$, and
$\rmm_\rma(0)\ge2N$. If we set $\vD\in\R^{N\times N}$ as
\[
\vD_{ij}=\langle\psi_i,\calL\psi_j\rangle,\quad i,j=1,\dots,n,
\]
then by the Fredholm alternative it will be the case that $\rmm_\rma(0)=2N$
if and only if $\vD$ is nonsingular: this will henceforth be assumed.

In $u_\lambda$ is a solution to \eref{e:11} for $\lambda\neq0$, then by the
Fredholm alternative it must be the case that for any $\phi\in\ker(\calL)$,
\[
0=-\langle\calJ^{-1}u_\lambda,\phi\rangle=\langle u_\lambda,\calJ^{-1}\phi\rangle
=\langle u_\lambda,\calL\psi\rangle.
\]
Consequently, the eigenvalue problem is not solved on all of $Y$, but is
instead solved on the $N$ co-dimensional constrained space $S^\perp$, where
$S=\Span\{\calL\psi_1,\dots,\calL\psi_N\}$. Thus, it will be important not to
calculate $\rmn(\calL)$, but instead $\rmn(\calL_{S^\perp})$, where
$L_{S^\perp}=P_{S^\perp}\calL P_{S^\perp}:S^\perp\mapsto S^\perp$, and
$P_{S^\perp}:X\mapsto S^\perp$ is the orthogonal projection. It was most
recently shown in \cite{kapitula:sif12} that
\begin{equation}\label{e:11a}
\rmn(\calL_{S^\perp})=\rmn(\calL)-\rmn(\vD);
\end{equation}
thus, the matrix $\vD$ also plays a significant role in determining the
number of negative eigenvalues for the constrained operator.

We are now ready to state the main result regarding the number of eigenvalues
for $\calJ\calL$ which have positive real part. Let
\begin{itemize}
\item $k_\rmr$: the total number of positive real-valued eigenvalues
    (including algebraic multiplicity)
\item $k_\rmc$: the total number of complex eigenvalues with positive real
    and imaginary part (including algebraic multiplicity)
\end{itemize}
Our assumptions on $\calJ$ and $\calL$ imply the four-fold eigenvalue
symmetry $\{\pm\lambda,\pm\overline{\lambda}\}$, so there will be
$k_\rmr+2k_\rmc$ eigenvalues with positive real part. There is one more set
of eigenvalues which we wish to count. First, for a self-adjoint operator
$\calT$ and a subspace $Z$ with basis $\{z_1,\dots,z_d\}$, let the Hermitian
matrix $\calT_Z\in\C^{d\times d}$ be given by
\[
(\calT_Z)_{ij}=\langle z_i,\calT z_j\rangle,\quad i,j=1,\dots,d.
\]
With this notation, for each nonzero eigenvalue $\lambda\in\rmi\R^+$ with
associated eigenspace $\E_\lambda$, let
\[
k_\rmi^-(\lambda)\coloneqq\rmn(\calL_{\E_\lambda}).
\]
The quantity $k_\rmi^-(\lambda)$ is known as the negative Krein index of the
eigenvalue, and if $k_\rmi^-(\lambda)\ge1$ the eigenvalue is said to have
negative Krein signature. Counting only those purely imaginary eigenvalues
with positive imaginary part, we say the total negative Krein index is given
by
\[
k_\rmi^-=\sum_{\lambda\in\sigma(\calJ\calL)\cap(\rmi\R^+\backslash\{0\})}k_\rmi^-(\lambda).
\]
Although we will not prove it here, it can easily be shown that
$k_\rmi^-(\overline{\lambda})=k_\rmi^-(\lambda)$ for
$\lambda\in\sigma(\calJ\calL)\cap\rmi\R$. Consequently, there will be
$2k_\rmi^-$ purely imaginary eigenvalues with negative Krein signature. The
Hamiltonian-Krein index is the weighted sum of these indices; namely;
\begin{equation}\label{e:12}
K_{\mathrm{Ham}}=k_\rmr+2k_\rmc+2k_\rmi^-.
\end{equation}
The first two terms in the index count the total number of eigenvalues with
positive real part, and the last term counts all those purely imaginary
eigenvalues with negative Krein signature. The major result is that the
Hamiltonian-Krein index is intimately related to the number of negative
eigenvalues of the constrained operator,
\begin{equation}\label{e:13}
K_{\mathrm{Ham}}=\rmn(\calL)-\rmn(\vD)
\end{equation}
\citep{haragus:ots08,pelinovsky:ilf05}. As a consequence of \eref{e:13} we
know that there is a finite and prescribed number of (potentially) unstable
eigenvalues, which, as we will see in subsequent sections, greatly
assists in a numerical search for these eigenvalues.

\begin{remark}
Additional implications of \eref{e:13} are:
\begin{enumerate}
\item if $K_{\mathrm{Ham}}$ is odd, then $k_\rmr\ge1$, so that the underlying wave is
spectrally unstable
\item if $K_{\mathrm{Ham}}=0$, then it is generically the case that the wave
is orbitally stable
\item if $K_{\mathrm{Ham}}$ is even, then the wave may be spectrally stable
    with $k_\rmr=k_\rmc=0$; however, the orbital stability of the wave is
    generally not known
\end{enumerate}
\end{remark}

\begin{remark}
The Krein signature has important implications beyond what is present in the
Hamiltonian-Krein index. If two purely imaginary eigenvalues collide, then
after the collision they can attain a nonzero real part (this is the
so-called Hamiltonian-Hopf bifurcation) if and only if they have opposite
signature. If the two eigenvalues have the same signature, they will simply
pass through each other. For a more detailed discussion see
\citet[Chapter~7.1]{kapitula:sad13}. Consequently, we can think of
eigenvalues having negative Krein signature as being potentially unstable
eigenvalues.
\end{remark}

\begin{remark}\label{r:23}
If the eigenvalue problem is canonical, i.e., of the form
\[
-\calL_+u=\lambda v,\quad \calL_-v=\lambda u,
\]
where $\calL_\pm$ are self-adjoint operators, then it is possible to derive a
lower bound on $k_\rmr$. For example, if $\dim[\ker(\calL_+)]=1$ and
$\dim[\ker(\calL_-)]=0$, then it is true that if
$|\rmn(\calL_+)-\rmn(\calL_-)|\ge2$, then $k_\rmr\ge1$
\citep{jones:ios88,grillakis:aot90}. This result will be discussed in further
detail in \autoref{s:4}.
\end{remark}

Before we can construct the Krein matrix associated with the general
Hamiltonian spectral problem \eref{e:11}, we must first find an equivalent
self-adjoint pencil. In addition, we must relate the Hamiltonian-Krein index
for the original problem to that of the pencil problem. First suppose that
$\Re\lambda>0$. Upon setting $v=\calJ^{-1}u$, and defining
\[
\calL_+=\calL,\quad\calL_-=-\calJ\calL\calJ,
\]
it is not difficult to see that solving \eref{e:11} is equivalent to solving
the canonical system
\begin{equation}\label{e:14}
\calL_+u=\lambda v,\quad \calL_-v=-\lambda u.
\end{equation}
We continue by writing down an equivalent eigenvalue problem for which the
operators involved no longer have a nontrivial kernel. Regarding
$\ker(\calL_\pm)$ we have
\[
\ker(\calL_+)=\ker(\calL),\quad
\ker(\calL_-)=\Span\{\calJ^{-1}\phi_1,\dots,\calJ^{-1}\phi_N\}=\calJ^{-1}\ker(\calL):
\]
the discussion in the previous subsection allows us to say that
$\ker(\calL_+)\perp\ker(\calL_-)$. Upon setting
$\Pi:X\mapsto[\ker(\calL_+)\oplus\ker(\calL_-)]^\perp$ to be the orthogonal
projection, it is known from \cite[Section~3]{kapitula:cev04} that for
nonzero eigenvalues \eref{e:14} is equivalent to the system
\begin{equation}\label{e:15}
-\Pi\calL_+\Pi u=\lambda v,\quad\Pi\calL_-\Pi v=\lambda u.
\end{equation}
Each of the operators $\Pi\calL_\pm\Pi$ are self-adjoint, and the assumption
that $\vD$ is nonsingular implies that each is also nonsingular on the range
of $\Pi$. This allows us to introduce the invertible self-adjoint operators
\begin{equation}\label{e:15a}
\calR\coloneqq\Pi\calL_+\Pi,\quad\calS^{-1}\coloneqq\Pi\calL_-\Pi,
\end{equation}
and note that $\calJ,\Pi$ being a bounded operators implies that from our
original assumptions on $\calL$ we have $\calR,\calS^{-1}:Y\mapsto X$. and
rewrite \eref{e:15} as the quadratic pencil
\[
(\calR+\lambda^2\calS)u=0,\quad u\in Y.
\]
In a similar fashion, if we initially assumed that $\lambda\in\rmi\R^+$, the
the equivalent pencil would be
\[
(\calR+\lambda^2\calS)[\Im u]=0,\quad \Re u=(\Im\lambda)\,\calJ\calS[\Im u].
\]

\begin{figure}[ht]
\begin{center}
\includegraphics{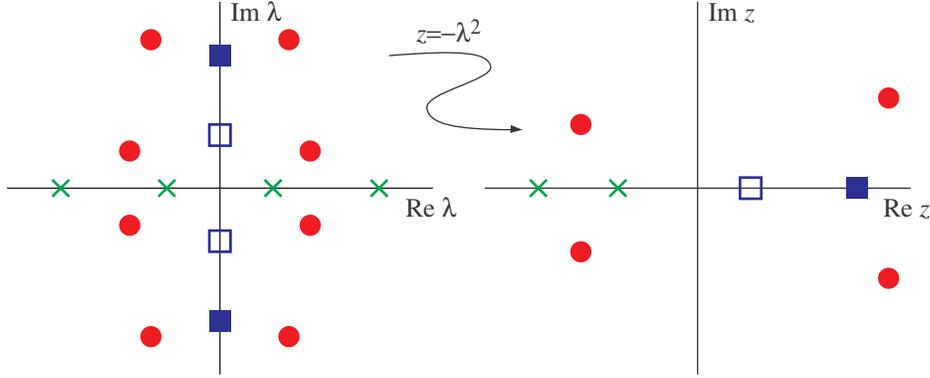}
\end{center}
\caption{(color online) Six sets of eigenvalues and their images under the map. The (red) circles denote two quads of complex
eigenvalues under the four-fold symmetry $\{\pm\lambda, \pm \overline{\lambda}\}$ and their images, $k_\rmc=2$. The (green) crosses denote
two pairs of real eigenvalues $\{\pm \lambda\}$ and their images on the negative real axis, $k_\rmr=2$. The (blue) boxes denote two pairs of purely imaginary
eigenvalues $\{\pm \lambda\}$ and their images on the positive real axis. The filled square has a positive Krein signature, while the
empty square has a negative Krein signature, so that $k_\rmi^-=1$.}
\label{f:SpectralPlaneMap}
\end{figure}

In conclusion, we have that solving the original eigenvalue problem
\eref{e:11} is equivalent to solving the linear pencil
\begin{equation}\label{e:16}
(\calR-z\calS)u=0,\quad
z\coloneqq-\lambda^2\quad(-\pi/2<\arg\lambda\le\pi/2),
\end{equation}
which is precisely the spectral problem that was studied by
\citet{grillakis:aot90}. The effect of the eigenvalue mapping is illustrated
in \autoref{f:SpectralPlaneMap}. Eigenvalues with positive real part and
nonzero imaginary part are mapped in a one-to-one fashion to eigenvalues with
nonzero imaginary part, and the four-fold symmetry is reduced to the
reflection symmetry $\{z,\overline{z}\}$. The system \eref{e:15} has an
unstable eigenvalue $\lambda$ with positive real part if and only if the
system \eref{e:16} has an eigenvalue $z$ with $z<0$ or with $\Im z\neq0$.

Let us conclude by computing the Hamiltonian-Krein index for the pencil
\eref{e:16}. First consider the purely imaginary eigenvalues. It is
straightforward to show that
\[
\calL_{\E_\lambda}=2\calR_{\E_z},
\]
where $z=-\lambda^2\in\R^+$, and $\E_z=\Im\E_\lambda$. Since
$\E_{\pm\lambda}=\Re\E_\lambda\pm\rmi\Im\E_\lambda$, it is consequently the
case that
\[
k_\rmi^-(z)=k_\rmi^-(\lambda)+k_\rmi^-(-\lambda)=2k_\rmi^-(\lambda),
\]
where for $z\in\R^+$ the Krein index is found by computing
\[
k_\rmi^-(z)=\rmn(\calR_{\E_z}).
\]
Thus, the total negative Krein index is the same for the pencil as it is for
the original problem. Now consider those eigenvalues for the original problem
with nonzero real part. Set $k_\rmr(z)$ to be the multiplicity of the
real-valued eigenvalue $z\in\R^+$, and let $k_\rmc(z)$ be the multiplicity of
the eigenvalue for $\Im z\neq0$ (it is clearly the case that
$k_\rmc(\overline{z})=k_\rmc(z)$). Since
\[
(\calR-(\pm\lambda)^2\calS)u_{\pm\lambda}=0,
\]
we clearly have that
\[
k_\rmr(z)=2k_\rmr(\lambda),\quad k_\rmc(z)=2k_\rmc(\lambda).
\]
Here we are using the notation that $k_\rmr(\lambda)$ is the multiplicity of
the positive real-valued eigenvalue for \eref{e:11}, and $k_\rmc(\lambda)$ is
the multiplicity of an eigenvalue with positive real and imaginary parts.
Upon summing over all of the eigenvalues, and using \eref{e:12}, we get the
new result that the Hamiltonian-Krein index for the original problem is
related to that of the pencil \eref{e:16} in the following manner:

\begin{lemma}\label{lem:kreinindex}
Consider the linear pencil \eref{e:16} as derived from the eigenvalue problem
\eref{e:11}. For the pencil let $k_\rmr$ denote the total number of negative
real-valued eigenvalues (counting multiplicity), $k_\rmc$ the total number of
eigenvalues with positive imaginary part (counting multiplicity), and
$k_\rmi^-$ the total negative Krein index of all the positive real-valued
eigenvalues, where the index for a single eigenvalue $z\in\R^+$ with
associated eigenspace $\E_z$ is given by
\[
k_\rmi^-(z)=\rmn(\calR_{\E_z}).
\]
Then with the Hamiltonian-Krein index given as in \eref{e:13}, the
eigenvalues for the pencil satisfy
\[
k_\rmr+2k_\rmc+2k_\rmi^-=2K_{\mathrm{Ham}}.
\]
\end{lemma}

\begin{remark}\label{r:25}
Abusing notation a bit, we will, e.g., denote $k_\rmr(z)$ as the total number
of negative real eigenvalues for the pencil, and $k_\rmr(\lambda)$ the total
number of positive real eigenvalues for the original eigenvalue problem. We
can then summarize the above discussion to say
\[
k_\rmr(z)=2k_\rmr(\lambda),\quad k_\rmc(z)=2k_\rmc(\lambda),\quad
k_\rmi^-(z)=2k_\rmi^-(\lambda),
\]
and that the indices are
\[
\begin{split}
\left(k_\rmr+2k_\rmc+2k_\rmi^-\right)(\lambda)&=K_{\mathrm{Ham}},\quad\,\,\,\,\,\mathrm{problem}\,\,\eref{e:11}\\
\left(k_\rmr+2k_\rmc+2k_\rmi^-\right)(z)&=2K_{\mathrm{Ham}},\quad\mathrm{problem}\,\,\eref{e:16}.
\end{split}
\]
\end{remark}

It will be convenient to relate the Hamiltonian-Krein index to
$\rmn(\calS^{-1})=\rmn(\calR)$ (the equality follows from the fact that
$\calJ$ has bounded inverse). As for the number of negative directions for
$\calS^{-1}$, we have
\begin{equation}\label{e:16a}
\rmn(\calS^{-1})=\rmn(-(\calJ\calL\calJ)_{\ker(\calL)^\perp})=
\rmn(\calL_{[\calJ^{-1}\ker(\calL)]^\perp})=
\rmn(\calL)-\rmn(\vD)=K_{\mathrm{Ham}}.
\end{equation}
The first equality follows from the definition of $\calS^{-1}$ and the fact
that constrained operator maps the subspace $\ker(\calL)^\perp$ to itself,
the second equality follows from a simple change of variables, the third
equality follows from \eref{e:11a}, and the fourth equality is from
\eref{e:13}. Thus, since the number of negative directions is invariant under
inversion, with respect to the pencil alone we can rewrite the conclusion of
\autoref{lem:kreinindex} as
\begin{equation}\label{e:17}
\left(k_\rmr+2k_\rmc+2k_\rmi^-\right)(z)=2\rmn(\calS).
\end{equation}
It can be concluded that if the underlying wave is orbitally stable, then
$\calS$ is positive definite; otherwise, the operator must be indefinite,
although it will necessarily have only a finite number of negative
directions.

\subsection{The Krein matrix}

We now turn to the problem of constructing a meromorphic matrix-valued
function, the Krein matrix, which has the property that it is singular
precisely for those values of $z$ which correspond to nonzero eigenvalues for
the pencil
\begin{equation}\label{e:21}
(\calR-z\calS)u=0
\end{equation}
where $\calR,\calS$ are defined in \eref{e:15a}. The Krein matrix was
introduced in \cite{kapitula:tks10}, and the interested reader should consult
that work for the details associated with the following discussion (also see
\cite[Section~3]{kapitula:sif12}). We construct the Krein matrix by
projecting off the finite-dimensional negative space of the operator $\calS$,
which as we have already seen in \eref{e:16a} has dimension
$K_{\mathrm{Ham}}$, and then using a Lyapunov-Schmidt reduction to compute an
equivalent eigenvalue problem.

Let $N(\calS)$ denote the $K_{\mathrm{Ham}}$-dimensional negative subspace of
$S$, and let $P:X\mapsto N(\calS)^\perp$ be the orthogonal projection. Define
the constrained operators
\[
\calR_2\coloneqq P\calR P,\quad\calS_+\coloneqq P\calS P.
\]
The operator $\calS_+$ is positive definite and symmetric, whereas from the
Index Theorem \cite[Theorem~2.1]{kapitula:sif12} the symmetric operator
$\calR_2$ satisfies
\[
\rmn(\calR_2)=\rmn(\calR)-\rmn(\calR^{-1}_{N(\calS)})=
K_{\mathrm{Ham}}-\rmn(\calR^{-1}_{N(\calS)}).
\]
Since $\calS_+$ is positive definite we can define the conjugated operator
\[
\widetilde{\calR}\coloneqq\calS_+^{-1/2}\calR_2\calS_+^{-1/2},
\]
and note that the invertibility of $\calS$ yields
\[
\rmn(\widetilde{\calR})=\rmn(\calR_2).
\]

Upon writing a potential eigenfunction to the linear pencil as
\[
u=\sum_{j=1}^{K_{\mathrm{Ham}}}c_js_j^-+s^+,\quad s^+\in N(\calS)^\perp,
\]
where $\{s_1,\dots,s_{K_{\mathrm{Ham}}}\}$ is an orthonormal spectral basis
for $N(\calS)$, and applying the projection $P$ to \eref{e:21}, the pencil
problem becomes
\[
(\calR_2-z\calS_2)s^++\sum_{j=1}^{K_{\mathrm{Ham}}}c_jP\calR s_j=0.
\]
On the other hand, upon applying $Q=\calI-P$ to \eref{e:21}, where $\calI$ is
the identity operator, and taking the inner product of the resulting equation
with $s_\ell$ for $\ell=1,\dots,K_{\mathrm{Ham}}$, yields the system of
equations
\[
\langle\calR s^+,s_\ell\rangle+
\sum_{j=1}^{K_{\mathrm{Ham}}}c_j\langle\calR s_j,s_\ell\rangle=
c_\ell z\lambda_\ell,
\]
where $\calS s_\ell=\lambda_\ell s_\ell$. Solving the first equation for
$s^+$ and plugging this result into the second equation yields the problem
\begin{equation}\label{e:18}
\vK(z)\vc=\vn,\quad\vK(z)\coloneqq\calR_{N(\calS)}-z\calS_{N(\calS)}-
(\widetilde{\calR}-z)^{-1}_{\calS_+^{-1/2}P\calR N(\calS)}.
\end{equation}
Here we are using the notation
\[
\calS_+^{-1/2}P\calR N(\calS)\coloneqq\{\calS_+^{-1/2}P\calR s^-:s^-\in N(\calS)\}.
\]
The matrix $\vK(z)$ is known as the \textsl{Krein matrix}.

We now relate the properties of the problem \eref{e:18} to those for the
original pencil \eref{e:21}. By construction $z$ is an eigenvalue for the
pencil \eref{e:16} with corresponding eigenfunction $u$ if and only if
$\vK(z)\vc=\vn$, where $\vc=(c_1,\dots,c_{K_{\mathrm{Ham}}})^\rmT$. If for a
particular eigenvalue $z$ it is true that $u\notin N(\calS)^\perp$, then it
is necessarily true that $\det\vK(z)=0$. On the other hand, if $u\in
N(\calS)^\perp$, then $\vc=\vn$, so it can be the case that
$\det\vK(z)\neq0$. Now, if $\Im z\neq0$, then it will always be the case that
$z$ is an eigenvalue if and only if $\det\vK(z)=0$. On the other hand, if
$z\in\R^-$ is an eigenvalue with $\det\vK(z)\neq0$, then it is the case that
for the Krein matrix constructed by projecting off of the negative directions
of $\calR$, say $\vK_{\calR}(z)$, we would necessarily have
$\det\vK_{\calR}(z)=0$. Finally, if $z\in\R^+$ is an eigenvalue with
$\det\vK(z)\neq0$, then the eigenvalue has positive Krein signature. In this
case the eigenvalue is realized as a removable singularity of the Krein
matrix, i.e., $z=z_\rmp$ is a pole of the Krein matrix for which the residue
is the zero matrix. If the eigenvalue has negative Krein signature, it will
necessarily be the case that $\det\vK(z)=0$.

In conclusion, we can use $\det\vK(z)$ as a meromorphic function whose zeros
correspond to eigenvalues. The (potential) singularities of the Krein matrix
arise through $(\widetilde{\calR}-z)^{-1}_{\calS_+^{-1/2}P\calR N(\calS)}$ at
the eigenvalues of the self-adjoint operator $\widetilde{\calR}$. In order to
use the Krein matrix to say something about the Krein signature of a
real positive eigenvalue, it will be helpful to factor the determinant as a
finite product. One can easily observe that $\vK(z)$ is symmetric for all
$z$, i.e., $\vK(z)^\rmT=\vK(z)$; in particular, it is Hermitian for $z\in\R$.
This allows us for $z\in\R$ to extract the $K_{\mathrm{Ham}}$ eigenvalues of
the Krein matrix, $r_j(z)$, hereafter called the \textsl{Krein eigenvalues},
in a meromorphic fashion. Thus, instead of finding eigenvalues by looking for
the zeros of the determinant of the Krein matrix, we can look for the zeros
of each individual Krein eigenvalue. There will be precisely
$K_{\mathrm{Ham}}$ of these Krein eigenvalues.

\begin{remark}
Although we will not pursue this line of thought herein, each Krein
eigenvalue can be thought of as a real meromorphic analogue of the Evans
function \citep{alexander:ati90}. Both Krein eigenvalues and the Evans
function detect eigenvalue for a linear eigenvalue problem through the zeros.
Until recently the Evans function was constructed solely via a dynamical
systems argument, which necessitated that the eigenvalue problem be
essentially in one space variable. Since the Krein matrix is constructed via
a functional analytic argument, the spatial dimension associated with the
eigenvalue problem is not relevant.
\end{remark}

\begin{figure}[ht]
\begin{center}
\includegraphics{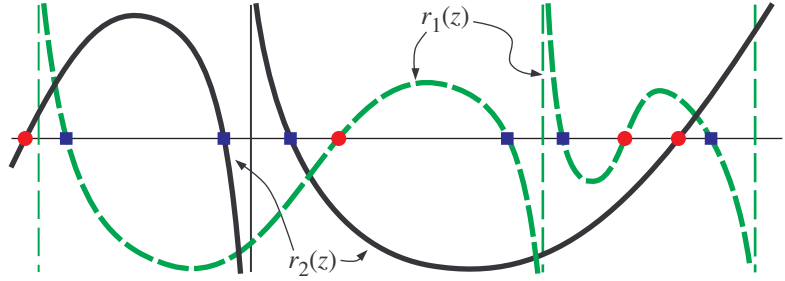}
\end{center}
\caption{(color online) A cartoon of the graphs of the Krein eigenvalues for $z\in\R^+$
in the case that $K_{\mathrm{Ham}}=2$. The Krein eigenvalue $r_1(z)$ is denoted by a thick (green)
dashed curve, and its vertical asymptotes are given by a thin (green) dashed curve. The Krein
eigenvalue $r_2(z)$ is denoted by a thick (black)
curve, and its vertical asymptotes are given by a thin (black) curve. The eigenvalues for the pencil
\eref{e:21} with positive Krein index are denoted by
(blue) squares, and those with negative index are shown as (red) circles.
} 
\label{f:TwoKrein}
\end{figure}

For real-valued $z$, the properties of the Krein eigenvalues are as follows.
First,
\[
\lim_{z\to-\infty}\frac{r_j(z)}{z}>0,
\]
so that each Krein eigenvalue is negative for large negative $z$. This
follows from the fact that $\calS_{N(\calS)}$ is a negative definite matrix.
Second, if $z_\rmp$ is a pole of the the Krein eigenvalue $r_j(z)$, it will
then be the case that
\[
\lim_{z\to z_\rmp^\pm}r_j(z)=\pm\infty.
\]
Furthermore, if $z^*$ is a simple eigenvalue of $\widetilde{\calR}$, so that
$z^*$ is a pole of the Krein matrix, then it will be the case that $z^*$ is a
simple pole for only one of the Krein eigenvalues. In other words, for all
but one of the Krein eigenvalues the pole $z^*$ of the Krein matrix is a
removable singularity. Finally, if $z_0$ is a simple positive zero of the
Krein eigenvalue $r_j(z)$, then it is true that
\[
k^-_\rmi(z_0)=-\sgn[r_j'(z_0)];
\]
in other words, the slope of the Krein eigenvalue at a zero gives definitive
information regarding the Krein index of the eigenvalue. This is the most
useful of the properties of the Krein eigenvalues, for it allows us
graphically locate those eigenvalues which have negative Krein signature.
This discussion is summarized in \autoref{f:TwoKrein}.

\begin{remark}
If the zero of a Krein eigenvalue is not simple, then for the eigenvalue in
question there will be an associated Jordan chain which has length equal to
the order of the zero. Furthermore, this situation can only arise upon the
collision of an (almost) equal number of purely imaginary eigenvalues with
positive and negative Krein signature (see
\citep[Section~2.2]{kapitula:tks10} for the details).
\end{remark}

\subsection{Summarizing remarks}\label{s:23}

Following \autoref{r:25} we know that for the original eigenvalue problem
\eref{e:11} the Hamiltonian-Krein index is
\[
\left(k_\rmr+2k_\rmc+2k_\rmi^-\right)(\lambda)=K_{\mathrm{Ham}},
\]
while for the pencil \eref{e:21} the index is
\[
\left(k_\rmr+2k_\rmc+2k_\rmi^-\right)(z)=2K_{\mathrm{Ham}}.
\]
The individual indices are related via
\[
k_\rmr(z)=2k_\rmr(\lambda),\quad k_\rmc(z)=2k_\rmc(\lambda),\quad
k_\rmi^-(z)=2k_\rmi^-(\lambda).
\]
For the original eigenvalue problem there will be an infinite number of
purely imaginary eigenvalues, all of which but a finite number will have
positive Krein index. The purely imaginary eigenvalues with negative Krein
signature can be determined by first constructing the Krein matrix of
\eref{e:18} for the pencil \eref{e:21}, and then plotting the resultant Krein
eigenvalues for $z\in\R$. In particular, the eigenvalues with negative
signature will correspond to those values of $z\in\R$ such that for some
$1\le j\le K_{\mathrm{Ham}}$,
\[
r_j(z)=0,\quad r_j'(z)>0.
\]
If the order of a zero of a Krein eigenvalue is two or higher, then a
collision of eigenvalues of opposite Krein signature has occurred.

\begin{remark}
The discussion in this section assumed that the original Hamiltonian
eigenvalue problem is not in the canonical form of \autoref{r:23}. If the
problem is in canonical form, adjustments must be made: this is discussed in
the application presented in \autoref{s:4}.
\end{remark}

\section{Application: spectral analysis for vortices of the GP
equation}\label{s:3}

We now wish to use the Krein matrix to identify the eigenvalues of
\eref{e:11} which have nonzero real part, or which are purely imaginary and have negative Krein signature. We intend to explore these spectral features
and confirm them against a full linear stability analysis
for an example of significant interest to recent experimental applications,
namely the study of a single vortex~\cite{dsh1}
and of a pair of vortices~\cite{bpa,dsh2} in two-dimensional
Bose-Einstein condensates as described by the GP Equation.
The Hamiltonian-Krein index of \eref{e:17} tells us how many of
these zeros for the pencil we need to find in order to fully capture all of the
(potentially) unstable eigenvalues. The real eigenvalues for the linearized problem correspond to negative real eigenvalues for the pencil, and the eigenvalues with nonzero real part correspond to eigenvalues with nonzero imaginary part for the pencil. Regarding the eigenvalues with negative signature, from the discussion of the previous section this means that we need to
identify the real and positive eigenvalues for the pencil \eref{e:16} for
which a Krein eigenvalue has a zero, and the slope of the curve at the zero
is positive.

The model under consideration for the case of pancake-shaped
Bose-Einstein condensates~\cite{emergent,dsh1,dsh2}
is the (2+1)-dimensional Gross-Pitaevskii equation
in the dimensionless form
\begin{equation}
\rmi\partial_t u = -\frac{1}{2}\Delta u+V(x,y)u+|u|^2u-\mu u.
\label{GPE_2d}
\end{equation}
Here, $u$ is the macroscopic wave function,
$V(x,y)=\Omega^2\left(x^2+y^2\right)/2$ is the external harmonic potential,
$\Omega$ is the frequency of the trap, $\mu$ is the chemical potential, and
$\Delta$ is the Laplace operator, i.e., $\Delta=\partial_{xx}+\partial_{yy}$.
For the problem discussed herein, we assume $\Omega=0.2$.

We begin by assuming that the steady-state problem is solved, and we will
denote that solution by $u_0(x,y)=U_0(x,y)+\rmi V_0(x,y)$. Here $U_0,V_0$ are
real-valued functions. Writing
\[
U_0(x,y)+\rmi V_0(x,y)=\rho(r,\theta)\rme^{\rmi\phi(r,\theta)},\quad\tan\theta=y/x,
\]
where $r^2=x^2+y^2$, the wave will have the property that
$\rho(r,\theta)\to0$ exponentially fast as $r\to+\infty$. As for the
associated eigenvalue problem, abusing notation a bit and writing
\[
u\mapsto U_0+\rmi V_0+\epsilon(u+\rmi v),\quad 0<\epsilon\ll1,
\]
we see that at $\calO(\epsilon)$ we have the linear problem
\[
\partial_t\vu=\calJ\calL\vu,\quad \vu=(u,v)^\rmT,
\]
where
\[
\calJ=\left(\begin{array}{rr}0 & 1 \\-1 & 0 \end{array}\right),\quad
\calL=\left(\begin{array}{cc}-\frac12\Delta-\mu+V(r)+3U_0^2+V_0^2&2U_0V_0\\
2U_0V_0&-\frac12\Delta-\mu+V(r)+U_0^2+3V_0^2\end{array}\right).
\]
The eigenvalue problem of the form \eref{e:11}, i.e.,
\begin{equation}\label{e:32}
\calJ\calL\vu=\lambda\vu,
\end{equation}
arises upon using the separation of variables ansatz
$\vu\mapsto\vu\rme^{\lambda t}$.

With respect to the standard inner-product on $L^2(\R^2)\times L^2(\R^2)$ the
operator $\calJ$ is boundedly invertible and skew-symmetric, while the
operator $\calL$ is self-adjoint. Furthermore, since the potential $V(r)$
grows quadratically, and the magnitude of the solution
$|u_0(x,y)|=\rho(r,\theta)$ decays exponentially fast as $r\to+\infty$, it is
the case that $\calL$ has a compact resolvent. In order to construct the
Krein matrix for the spectral problem we first follow \eref{e:14} and set
\[
\calL_+=\calL,\quad\calL_-=-\calJ\calL\calJ.
\]
The construction of the Krein matrix requires that we consider the spectral
problem on the space perpendicular to the kernel of both operators. The fact
that solutions to \eref{GPE_2d} are invariant under $u(x,y)\mapsto
u(x,y)\rme^{\rmi\phi}$ and the spatial rotation $u(x,y)\mapsto
u(x\cos\theta-y\sin\theta,x\sin\theta+y\cos\theta)$ means that (generically)
the kernel of $\calL$ will be two-dimensional, so that (generically)
\begin{equation}\label{e:L-ker}
\begin{split}
\ker(\calL)=\ker(\calL_+)&=\Span\{\left(\begin{array}{r}-V_0\\U_0\end{array}\right),
\left(\begin{array}{c}(\partial_\theta\rho)\cos\phi-(\partial_\theta\phi)V_0\\
(\partial_\theta\rho)\sin\phi+(\partial_\theta\phi)U_0\end{array}\right)\}\\
\ker(-\calJ\calL\calJ)=\ker(\calL_-)&=\Span\{\left(\begin{array}{c}U_0\\V_0\end{array}\right),
\left(\begin{array}{r}(\partial_\theta\rho)\sin\phi+(\partial_\theta\phi)U_0\\
-(\partial_\theta\rho)\cos\phi+(\partial_\theta\phi)V_0\end{array}\right)\}.
\end{split}
\end{equation}
If the solution has a radially symmetric density, i.e., $\rho=\rho(r)$, then
the dimension of each kernel is (generically) one with
\[
\ker(\calL_+)=\Span\{\left(\begin{array}{r}-V_0\\U_0\end{array}\right)\},\quad
\ker(\calL_-)=\Span\{\left(\begin{array}{c}U_0\\V_0\end{array}\right)\}.
\]
With this information at hand, and setting $\Pi:L^2(\R^2)\times
L^2(\R^2)\mapsto[\ker(\calL_+)\oplus\ker(\calL_-)]^\perp$ to the the
orthogonal projection, we can now compute the restricted operators
\[
\calR=\Pi\calL_+\Pi,\quad\calS^{-1}=\Pi\calL_-\Pi
\]
to create the pencil
\[
(\calR-z\calS)\vu=\vn,\quad z=-\lambda^2.
\]
The Krein matrix $\vK(z)\in\C^{\rmn(\calS)\times\rmn(\calS)}$ (see
\eref{e:18}) will be created from this pencil using the algorithm described
in the previous section.

\subsection{Single vortex state}

\begin{figure}
\begin{center}
\includegraphics[width=9cm]{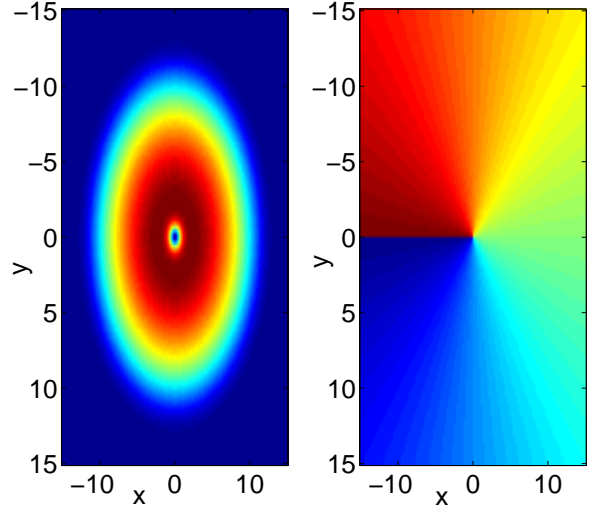}
\end{center}
\caption{(color online) The single vortex state for a trap strength $\Omega=0.2$ and chemical potential
$\mu=3$. The left panel
shows the contour plot of the density, and the right 
panel is the phase plot for the
wave function.
In this case, the total number of grid points is $n=120$, and
the spatial step size is $\Delta x=0.25$.
}
\label{single_vortex_density_phase_lsa}
\end{figure}

\begin{figure}
\begin{center}
\includegraphics[width=9cm]{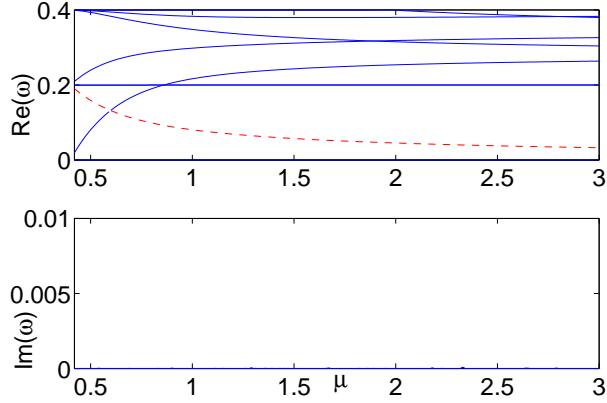}
\end{center}
\caption{(color online) The real and imaginary parts of the eigenfrequencies defined by
$\omega=\rmi\lambda$ for vortex dipole state when $n=120$ and
$\Delta x=0.25$ as a function of the chemical potential $\mu$. Since
the (red) dashed curve
emanating from $\omega=0.2$ corresponds to an eigenvalue with
negative Krein signature, the fact that $K_{\Ham}=2$ means that
there are no other eigenvalues with positive real part (see \eref{e:kvortex2}). The wave is spectrally
stable for all considered values of the chemical potential. The eigenvalue which has negative
Krein signature is highlighted as the (red) dashed line.
Larger (in magnitude) eigenvalues along the imaginary axis have a positive
Krein signature and do not lead to instabilities.}
\label{single_vortex_spectrum}
\end{figure}

Here we assume that the solution is a vortex of charge one. The solution is
of the form
\[
u_0(x,y)=\rho(r)\rme^{\rmi\theta},
\]
where the density profile satisfies $\rho(0)=0$, and the phase rotates by
$2\pi$ around the vortex core, justifying the topological charge of the
structure~\cite{emergent,kollar:sso12}. The density and phase profiles for
the single vortex state are shown in
\autoref{single_vortex_density_phase_lsa}, and the corresponding
linearization spectrum is shown in \autoref{single_vortex_spectrum}.
The absolute value of this configuration is radially symmetric
(a feature not shared by the vortex dipole state considered below),
while its decay for $r \rightarrow \infty$ is dictated by the
underlying linear problem being exponential for small
$\mu$~\cite{kapitula:rmw07} and resembling an inverted parabola for
large $\mu$~\cite{peli4} (the latter decay features are also shared
by the vortex dipoles below).
The
dependence of the relevant eigenvalues as a function of the canonical
parameter of the system, namely the chemical potential $\mu$ associated with
the number of atoms in the condensate, has been quantified previously; see
e.g.~\cite{kollar:sso12,middel10}. We will confirm these findings via the
Krein matrix, and showcase particular examples for a few representative
values of $\mu$.
As we saw above, the kernels of the operators $\calL_\pm$ each have dimension
one. For small amplitude vortices where the nonlinear interactions are
(almost) negligible it can be shown via an analysis similar to that presented
in \citep{kapitula:rmw07} that the Hamiltonian-Krein index satisfies
$K_{\Ham}=2$. Following the discussion in \autoref{s:23} we then know that
for the pencil
\begin{equation}\label{e:kvortex}
\left(k_\rmr+2k_\rmc+2k_\rmi^-\right)(z)=4,
\end{equation}
while for the original problem
\begin{equation}\label{e:kvortex2}
\left(k_\rmr+2k_\rmc+2k_\rmi^-\right)(\lambda)=2.
\end{equation}
Thus, if for the pencil there are two positive real zeros of the Krein
eigenvalues which correspond to an eigenvalue with negative Krein index, the
rest of the spectrum for the pencil must be positive and purely real. In
other words, if for the original eigenvalue problem there is a purely
imaginary eigenvalue with negative Krein index, the rest of the spectrum must
be purely imaginary with positive Krein index. Consequently, once we
numerically find one purely imaginary eigenvalue with negative Krein
signature, or one set of eigenvalues with nonzero real part which satisfy the
Hamiltonian eigenvalue symmetry, we need not search for any additional
unstable eigenvalues. They simply do not exist.

Regarding the numerical solution and analysis of the problem, we first
discretize the differential operator via a centered finite difference scheme
using $n$ data points and a spatial interval of $\Delta x$. The single vortex
state - indeed, any stationary solution - is found by applying Newton's
method to the discretized problem, treating it as a two-dimensional boundary
value problem and starting from a suitably proximal to it initial guess. As
for the spectral problem, the smallest (in norm) eigenvalues are computed
using the MATLAB function ``eigs" on the discretized version of the operator
$\calJ\calL$; i.e., we construct the linearization operator on the same
domain as used for the fixed point iteration and utilize standard routines
(such as an implicitly restarted Arnoldi method within MATLAB) to compute
some of the smallest magnitude eigenvalues of the corresponding spectral
problem. This provides us with the linear stability results that will be
compared with the Krein matrix ones in the figures that follow. It is worth
noting here that these spectral linearization results can only be obtained by
taking advantage of
the sparse structure of the underlying discretization matrix.
Should the full eigenvalue spectrum of the linearization problem be sought in
this highly-demanding two-dimensional computation, MATLAB's routine ``eig''
would have been unable to produce the corresponding numerical results.

\begin{figure}
\begin{center}
\includegraphics[width=8cm]{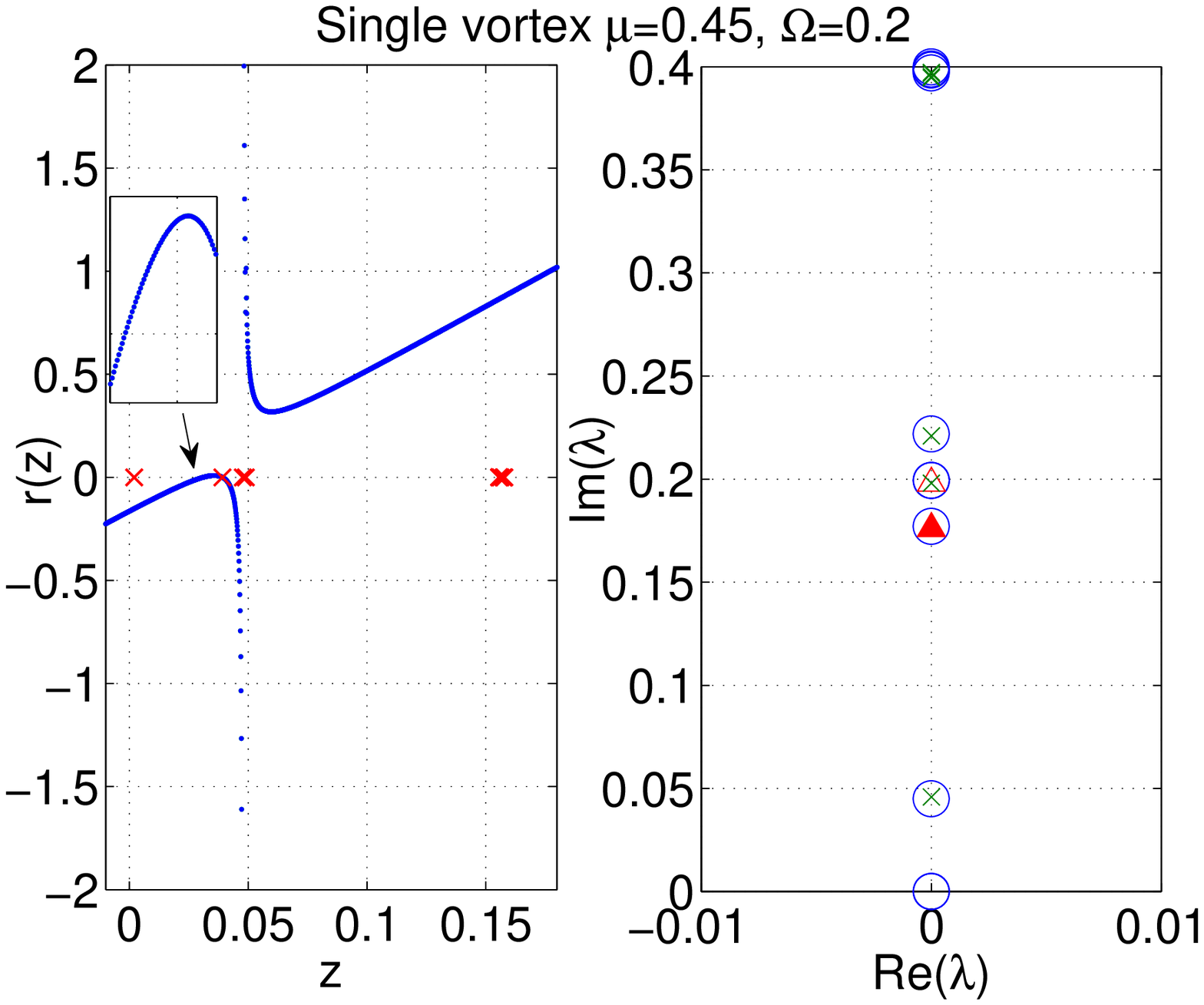}
\caption{(color online) The numerically generated spectral plot for $n=48,\,\mu=0.45,\,\Delta x=0.45$.
In the left panel the left plot is the numerically generated plot of the $K_{\Ham}=2$ Krein
eigenvalues. For this problem the Krein matrix is a meromorphic multiple of the identity; hence, the
two Krein eigenvalues coincide. The (red) crosses are the poles of the Krein matrix. If the pole
is removable (e.g., for $z\sim0.155$), then it corresponds to an eigenvalue with positive Krein
signature. Here we see two positive real zeros of the Krein eigenvalues for which the functions
have positive slope. Thus, for the pencil $k_\rmi^-(z)=2$ with $k_\rmr(z)=k_\rmc(z)=0$, which means that
for the original problem \eref{e:32} $k_\rmi^-(\lambda)=1$ with $k_\rmr(\lambda)=k_\rmc(\lambda)=0$. The wave is spectrally
stable, but is not a (local) minimizer for the constrained Hamiltonian. In the right figure the eigenvalues for $\calJ\calL$ are
denoted by (blue) circles. The (green) crosses represent the poles of the Krein matrix,
and the (red) triangles are the eigenvalues of $\calJ\calL$ which are realized as zeros
of the Krein eigenvalues. The purely imaginary eigenvalue with negative Krein signature
is shown as a \textit{filled} (red) triangle.
}
\label{comparison_lsa_full_spectrum_n_48_dx_05_mu_045}
\end{center}
\end{figure}

\begin{figure}
\begin{center}
\includegraphics[width=8cm]{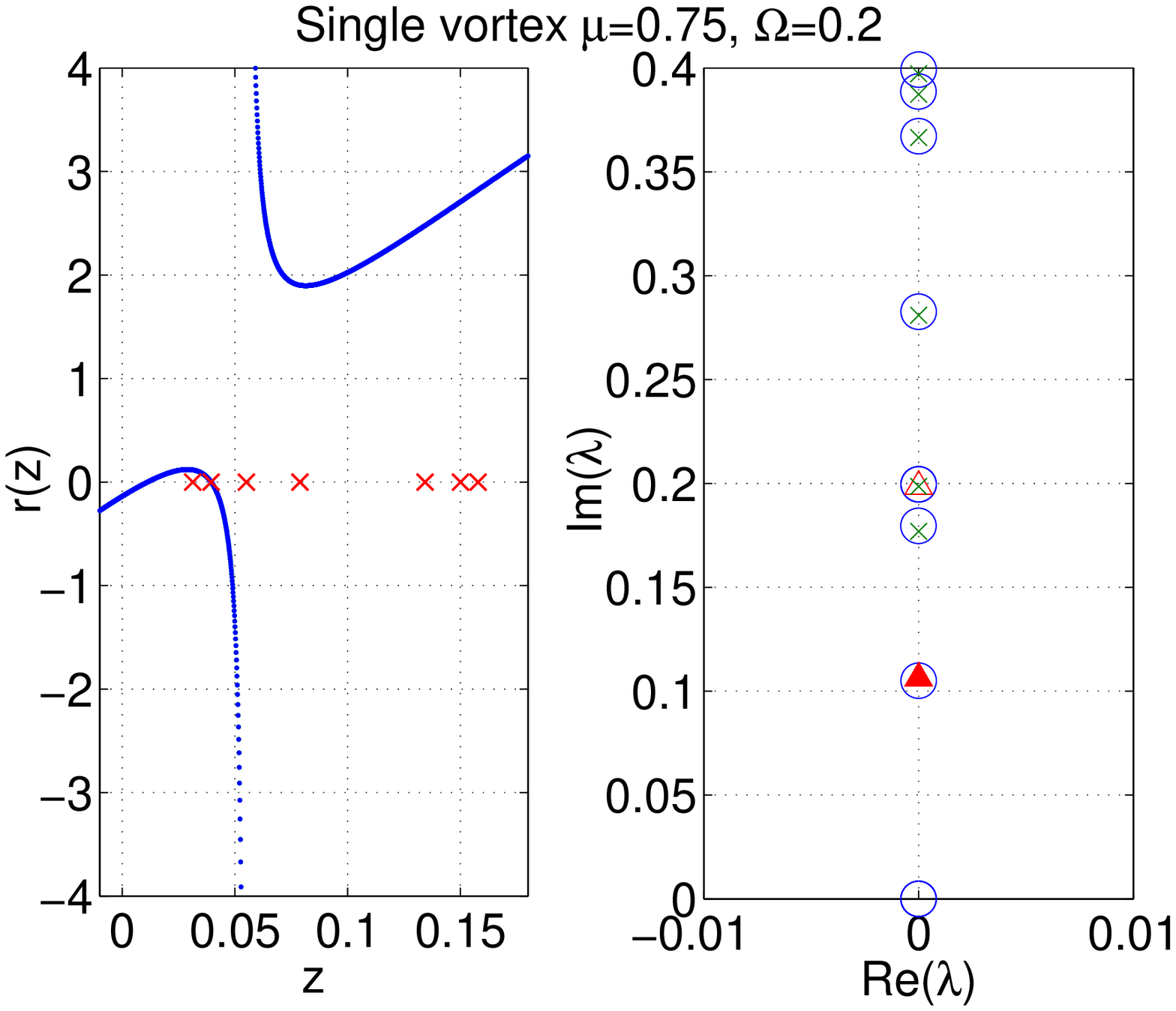}
\caption{(color online) The numerically generated spectral plot for $n=48,\,\mu=0.75,\,\rmd x=0.45$.
In the left panel the left plot is the numerically generated plot of the $K_{\Ham}=2$ Krein
eigenvalues. For this problem the Krein matrix is a meromorphic multiple of the identity; hence, the
two Krein eigenvalues coincide. The (red) crosses are the poles of the Krein matrix, and the
removable singularities (e.g., $z\sim0.08$) correspond to eigenvalues with positive Krein
signature. Here we again see two positive real zeros of the Krein eigenvalues for which the functions
have positive slope. The wave is then spectrally stable with $k_\rmi^-(\lambda)=1$
for the problem \eref{e:32}. The notation used in the right figure is similar to that in
\autoref{comparison_lsa_full_spectrum_n_48_dx_05_mu_045}.}
\label{comparison_lsa_full_spectrum_n_48_dx_058_mu_075}
\end{center}
\end{figure}

We now turn to a direct comparison of the numerical results for eigenvalues
obtained from the linear stability analysis
(which has partially been presented in earlier publications,
see e.g.~\cite{middel10})
and from the Krein matrix, presented for the first time for vortex
patterns of the GP equation herein.
Here,
we have confirmed the above findings with $\mu=0.45$ in
\autoref{comparison_lsa_full_spectrum_n_48_dx_05_mu_045}, and $\mu=0.75$ in
\autoref{comparison_lsa_full_spectrum_n_48_dx_058_mu_075}. In both of these
figures a spatial discretization spacing of $\Delta x=0.5$ was used. Similar
results were found with smaller values of $\Delta x$. In the right panel of
each figure the spectrum of $\calJ\calL$ was computed for both the original
formulation of \eref{e:32} and the corresponding pencil formulation
\eref{e:16}, and the results of the two were found to agree within the
accuracy of the eigenvalue solver. These spectral results were then compared
with the prediction of the Krein matrix in the right panel. The agreement is
excellent.

Unexpectedly, it turns out to be the case that for the problem at hand the
Krein matrix is a meromorphic multiple of the identity; hence, the two Krein
eigenvalues coincide. In both figures it is clear that there are then two
zeros of the Krein eigenvalues at which the slope of the curve is positive:
this corresponds to an purely imaginary eigenvalue with negative Krein
signature. As stipulated by Hamiltonian-Krein index in \eref{e:kvortex}, the
rest of the spectrum must be purely imaginary, and this is indeed seen to be
the case.

For the spectrum there is not only the eigenvalue at $\lambda=0$ which is due
to the phase U$(1)$ gauge invariance of the model, there is also a double
eigenvalue at $\lambda=\pm 0.2\rmi$. The latter frequency of double
multiplicity is the so-called dipolar mode of the condensate and pertains to
a symmetry (oscillation of the entire condensate cloud in the $x-$ or $y-$
direction with the trap frequency), and is hence invariant with respect to
variations in $\mu$. One of these modes pertains to a pole of the Krein
matrix, while the other is a zero of a Krein eigenvalue.
The eigenvalue with negative Krein signature always lies between the origin
and this double pair and is known to tend to the origin as the chemical
potential $\mu$ increases~\cite{middel10,kollar:sso12}. Since the eigenvalue
has negative signature, as predicted by the theory it is realized as a zero
of a Krein eigenvalue.

To showcase the relevance of this pole, the residues of the poles of the
Krein matrix are computed. If the numerically computed residue of the pole is
of $\calO(10^{-12})$, then we say that the pole is removable, and hence it
corresponds to an eigenvalue with positive Krein signature. The residue is
computed via the numerical integration
\[
\res(\vK(z),z_{\rmp})=
\frac1{2\pi\rmi}\oint_C \vK(z)\,\rmd z\sim\frac1{2\pi\rmi}\sum_{i=1}^{n}\vK(z_{\rmp,i})\Delta z_{\rmp,i}.
\]
Here $z_{\rmp}$ is the relevant pole of the Krein matrix, $n$ is the number
of integration points on the simple positively oriented closed contour $C$
which surrounds the pole (here the contour is without loss of generality
chosen to be a square centered on the pole), $z_{\rmp,i}$ is the point on the
integration contour around the pole, and $\Delta z_{\rmp,i}$ is the segment
on the integration path.

When the chemical potential $\mu=0.45$ the first relevant pole we choose is
located at $z_{\rmp}=0.0021$. For the increment of $|\Delta
z_{\rmp,i}|=5*10^{-6}$ it is seen that
\[
\oint_C \vK(z)\,\rmd z=\left( \begin{array}{ccc}
-8.4*10^{-12}-\rmi2.1*10^{-14} & -4.5*10^{-22}-\rmi9.5*10^{-20} \\
-2.8*10^{-21}-\rmi1.1*10^{-19} & -8.4*10^{-12}-\rmi2.1*10^{-14} \end{array} \right),
\]
so that $|\res(\vK(z),0.0021)|=\calO(10^{-12})$. The simple pole is
removable, and corresponds to a purely imaginary eigenvalue (the first red
cross above zero in
\autoref{comparison_lsa_full_spectrum_n_48_dx_05_mu_045}). The fact that the
singularity is removable is evidenced by the fact that neither of the Krein
eigenvalues has a singularity at the point. Now consider the pole at
$z_{\rmp}=0.0478$. It is seen that
\[
\oint_C \vK(z)\,\rmd z=\left( \begin{array}{ccc}
-\rmi0.0057 & -8.6*10^{-18}-\rmi3.0*10^{-15} \\
-5.5*10^{-18}-\rmi2.2*10^{-15} & -\rmi0.0057 \end{array} \right),
\]
so that $|\res(\vK(z),0.0478)|=\calO(10^{-3})$, which is nonzero by our
criterion. Alternatively, we see that the pole is not removable because the
Krein eigenvalues have a singularity at that point. Since the singularity is
not removable, this point does not correspond to an eigenvalue.

\begin{figure}
\begin{center}
\includegraphics[width=9cm]{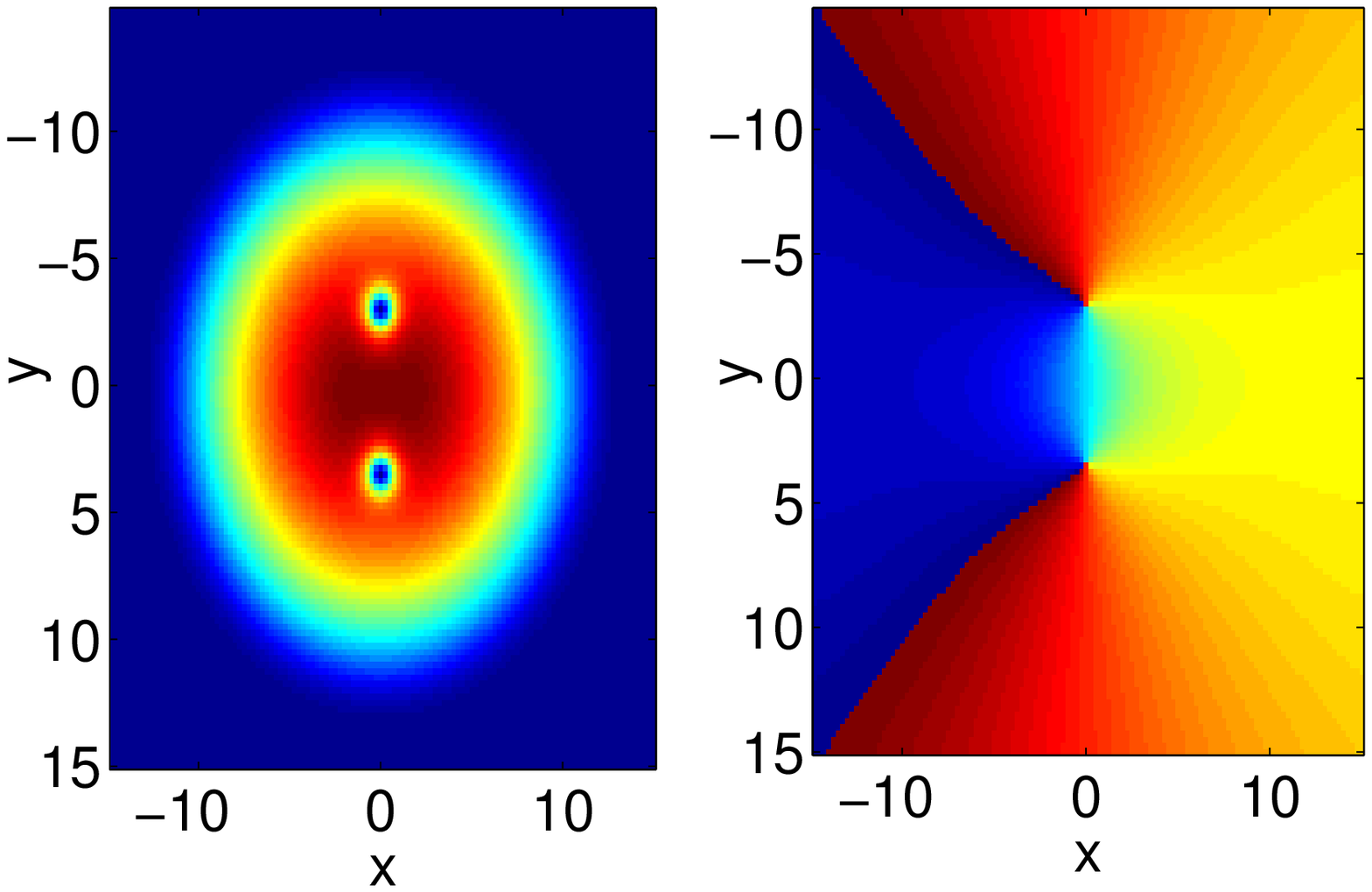}
\end{center}
\caption{(color online) The vortex dipole state for a trap strength $\Omega=0.2$ and chemical potential
$\mu=3$. The left panel
shows the contour plot of the density, and the right 
panel is the phase plot for the
wave function.
In this case, the total number of grid points is $n=120$, and
the spatial step size is $\rmd x=0.25$. 
}
\label{vortex_dipole_density_phase_lsa}
\end{figure}

\subsection{Vortex dipole state}
\begin{figure}
\begin{center}
\includegraphics[width=9cm]{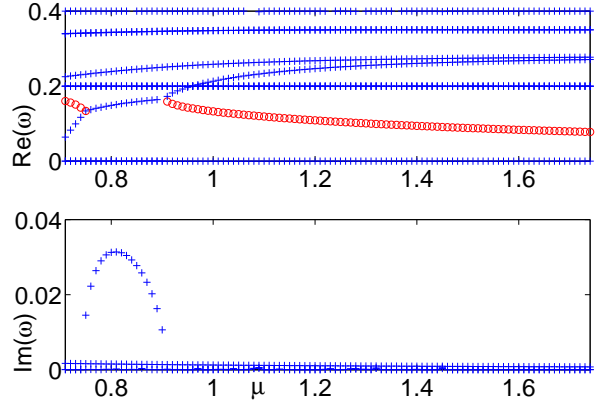}
\end{center}
\caption{(color online) The real and imaginary parts of the eigenfrequencies defined by
$\omega=\rmi\lambda$ for vortex dipole state when $n=480$ and
$\rmd x=0.0625$ as a function of the chemical potential $\mu$. The wave is spectrally
unstable with $k_\rmc(\lambda)=1,\,(k_\rmc(z)=2)$ when $\Im\omega>0$. When $k_\rmc(\lambda)=1$
we know by the Hamiltonian-Krein index for the wave that there are no other unstable eigenvalues.
When $k_\rmc(\lambda)=k_\rmr(\lambda)=0$ in the figure, the eigenvalue associated with
(red) circles has negative Krein signature. By the Hamiltonian-Krein index we then know that there are no other
non-plotted eigenvalues with positive real part, so that in this case the wave is spectrally stable.}
\label{vortex_dipole_density_phase_lsa_n_240_480}
\end{figure}

\begin{figure}
\begin{center}
\includegraphics[width=9cm]{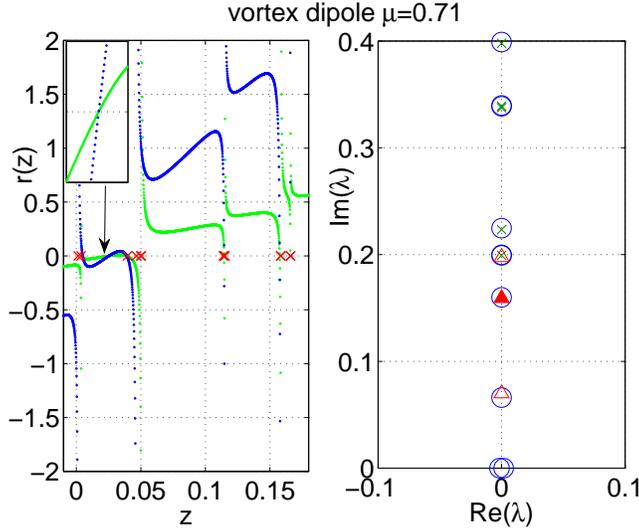}
\end{center}
\caption{(color online) The numerically generated spectral plot for the vortex dipole when $\Omega=0.2$ and
$\mu=0.71$. The left panel is the plot of the two Krein eigenvalues:
the (red) crosses represent the poles of the Krein matrix, and the removable singularities
correspond to eigenvalues with positive Krein signature. Note that $k_\rmi^-(z)=2$ for the pencil, and
that as expected the zeros of the Krein eigenvalues which correspond to eigenvalues with negative Krein
signature coincide. In the right panel the (blue) circles are the eigenvalues for $\calJ\calL$, the (red)
triangles are the eigenvalues which correspond to zeros of the Krein eigenvalues, and the (green)
crosses are the eigenvalues which correspond to removable singularities of the Krein matrix - these
eigenvalues have positive Krein signature. The purely imaginary eigenvalue with negative Krein signature
is shown as a \textit{filled} (red) triangle. The labeling of eigenvalues
in the right panel is the same as for \autoref{comparison_lsa_full_spectrum_n_48_dx_05_mu_045}.}
\label{vortex_dipole_eigen_krein_lsa_mu_071}
\end{figure}

We now turn to a spectral analysis for the vortex dipole state, which is a
stationary vortex-antivortex state (see
\autoref{vortex_dipole_density_phase_lsa} for a typical example of its
density and phase). 
When $\Omega=0.2$ the vortex dipole state exists for $\mu>0.68$.
It is interesting to note here that, as shown in~\cite{kom0,middel10}, such
states do not exist at the linear limit, but only bifurcate through a
supercritical pitchfork (symmetry-breaking) event at a critical point from
the dark soliton, which corresponds to the real-valued (first) excited state
of the two-dimensional harmonic oscillator. Since the density is not radially
symmetric, it will (generically) be the case that $\dim[\ker(\calL_\pm)]=2$.
Furthermore, it is seen numerically that $\rmn(\calS)=K_{\Ham}=2$, so that by
\autoref{lem:kreinindex} we have the same index count for the pencil as
\eref{e:kvortex}; namely,
\begin{equation}\label{e:kvortex}
\left(k_\rmr+2k_\rmc+2k_\rmi^-\right)(z)=4\quad\Leftrightarrow\quad
\left(k_\rmr+2k_\rmc+2k_\rmi^-\right)(\lambda)=2.
\end{equation}

As in the previous example, there will be two Krein eigenvalues to be
plotted. Unlike the previous example, the Krein matrix will not be a
meromorphic multiple of the identity; hence, the Krein eigenvalues will not
generically overlap. If there are two positive real zeros of the Krein
eigenvalues for which the curves have positive slope, then the spectrum for
the pencil will be positive and purely real with
$k_\rmi^-(z)=2\,(k_\rmi^-(\lambda)=1)$. Otherwise the underlying wave will be
unstable. As we will see, in our examples this instability will arise when
$k_\rmc(z)=2\,(k_\rmc(\lambda)=1)$. It is important to note here that for the
pencil it must be the case that if $k_\rmi^-(z)=2$, then each Krein
eigenvalue has a positive zero at precisely the same point. This fact is a
consequence of the relationship of the spectrum between the original
eigenvalue problem and the pencil; in particular, the fact that the spectrum
of the pencil ``doubles up" the nonzero spectrum for the original problem.

\begin{figure}
\begin{center}
\includegraphics[width=9cm]{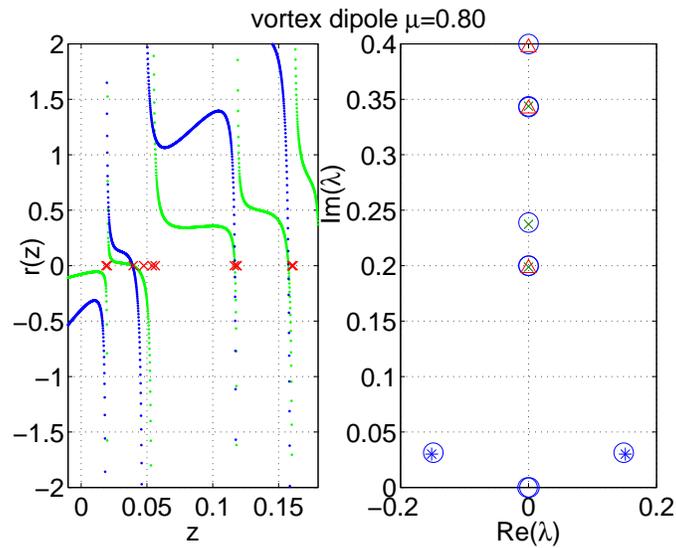}
\end{center}
\caption{(color online) The numerically generated spectral plot for the vortex dipole when
$\Omega=0.2$ and $\mu=0.80$. The left panel is the plot of the two Krein
eigenvalues: the (red) crosses represent the poles of the Krein matrix, and
the removable singularities correspond to eigenvalues with positive Krein
signature. Since there are no positive real zeros with positive slope, it
must be the case that for the pencil $k_\rmc(z)=2$. In the right panel the
(blue) circles are the eigenvalues for $\calJ\calL$, the (red) triangles are the
eigenvalues which correspond to zeros of the Krein eigenvalues, and the (green)
crosses are the eigenvalues which correspond to removable singularities of
the Krein matrix - these eigenvalues have positive Krein signature. The labeling of eigenvalues
in the right panel is the same as for \autoref{comparison_lsa_full_spectrum_n_48_dx_05_mu_045}.
The only difference is that the (blue) stars represent the
complex eigenvalues with nonzero real part as found by the zeros of the Krein eigenvalues. }
\label{vortex_dipole_eigen_krein_lsa_compare_mu_080_including_complex eigen}
\end{figure}

The spectrum once again features the twofold degenerate dipolar modes which
are associated with the oscillation with the trap frequency of the whole
condensate in the $x-$ and $y-$ directions. In addition to these modes, there
exists a negative Krein signature mode, which in this case collides with the
mode departing from zero. This, in turn, leads to the formation of an
interval of $\mu$-values where the spectrum possesses
 complex eigenfrequencies associated with oscillatory
instability. Therefore, when studying the computation of the Krein matrix of
the vortex dipole state and comparing its eigenvalues against the
linearization analysis, we select two cases, namely $\mu=0.71$ and
$\mu=0.80$. As we see in \autoref{vortex_dipole_density_phase_lsa_n_240_480},
the former is before the collision of the two modes (i.e.,
$k_\rmc(\lambda)=0$), and the latter is after the collision has occurred
(i.e., $k_\rmc(\lambda)=1$). For larger values of $\mu$ the complex-valued
eigenvalues return to the imaginary axis.

When $\mu=0.71$ we show the computation of the Krein eigenvalues (left panel)
and the linearization spectrum (right panel) in
\autoref{vortex_dipole_eigen_krein_lsa_mu_071}. In the right panel it can be
observed that the zeros of the Krein eigenvalues are consistent with the
eigenvalues found via the linear stability analysis, except for the real
eigenvalues, which are $\calO(10^{-3})$ and are the numerical approximation
of the known zero eigenvalues. Regarding the left panel it is seen from the
inset that $k_\rmi^-(z)\ge2$. From \eref{e:kvortex} we know that the
Hamiltonian-Krein index for the pencil is
\[
\left(k_\rmr+2k_\rmc+2k_\rmi^-\right)(z)=4;
\]
hence, for the pencil it is the case that $k_\rmi^-(z)=2$ with
$k_\rmr(z)=k_\rmc(z)=0$. For the original problem it is then true that
$k_\rmi^-(\lambda)=1$ with $k_\rmr(\lambda)=k_\rmc(\lambda)=0$. The
eigenvalue with negative Krein signature is denoted by a (red) filled
triangle in the right panel. As a consequence of the index theory we know
that all other eigenvalues must be purely imaginary with positive Krein
signature.

\begin{figure}
\begin{center}
\includegraphics[width=9cm]{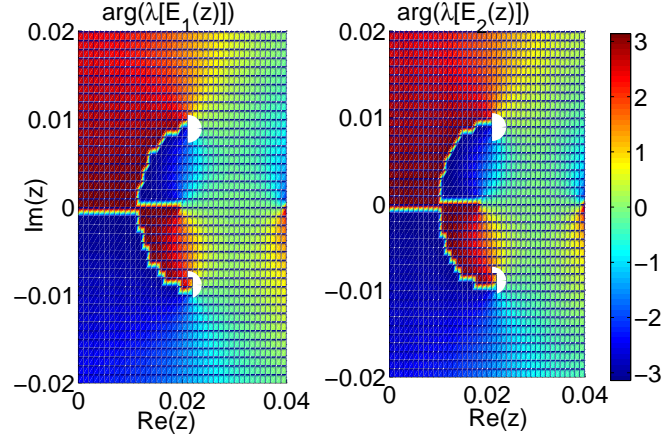}
\end{center}
\caption{(color online) Phase plot for the Krein eigenvalues when $\mu=0.80$. The colors
represent the argument of the Krein eigenvalues,
$\arg(\lambda(E_1(z)))$ and
$\arg(\lambda(E_2(z)))$. The two white spots on each of the two panels represent
the complex eigenvalues, corresponding to $z=0.021\pm\rmi0.09$.}
\label{vortex_dipole_domain_coloring}
\center
\end{figure}

When $\mu=0.80$ we show the computation of the Krein eigenvalues (left panel)
and the linearization spectrum (right panel) in
\autoref{vortex_dipole_eigen_krein_lsa_compare_mu_080_including_complex
eigen}. Again, the zeros of the Krein eigenvalues are consistent with the
eigenvalues found via the linear stability analysis except for the real
eigenvalues, which are $\calO(10^{-3})$ and are the numerical approximation
of the known zero eigenvalues. From the right panel we see that
$k_\rmc(\lambda)\ge1$. Upon using \eref{e:kvortex} we then have that
$k_\rmc(\lambda)=1\,(k_\rmc(z)=2)$ with
$k_\rmr(\lambda)=k_\rmi^-(\lambda)=0$. As we see in the left panel, as
expected the Krein eigenvalues have no negative zeros, and no positive zeros
with positive slope; in other words, all of the zeros of the Krein
eigenvalues correspond to purely imaginary eigenvalues with positive Krein
signature. In conclusion, as a consequence of the index theory we know that
except for the one quartet of simple eigenvalues with nonzero real and
imaginary parts, all other eigenvalues must be purely imaginary with positive
Krein signature. The unstable eigenvalues cannot be detected via the graph of
the Krein eigenvalues in the left panel of
\autoref{vortex_dipole_eigen_krein_lsa_compare_mu_080_including_complex
eigen}. However, they can be found by providing a phase plot for each Krein
eigenvalue, which is done in \autoref{vortex_dipole_domain_coloring} when
$\mu=0.80$. The axes denote the complex $z$-plane, and the colorbar
corresponds to the phase of the eigenvalues of the Krein matrix. The points
where the phase becomes singular correspond to the location of the complex
eigenvalues, and are labeled by the white spots. It is clear from a standard
winding number argument that each zero of the Krein eigenvalue is simple,
which is a verification of the fact that $k_\rmc(z)=2$.

\section{Application: spectral analysis for multi-solitons of the GP
equation}\label{s:4}

The model under consideration is the (1+1)-dimensional Gross-Pitaevskii
equation in the dimensionless form
\begin{equation}
\rmi\partial_t u = -\frac{1}{2}\partial_{xx} u+V(x)u+|u|^2u-\mu u,
\label{GPE_1d}
\end{equation}
where now $V(x)=\Omega^2x^2/2$ is the external harmonic potential. When doing
numerical computations for this problem we assume $\Omega=1$. For this
problem we are interested in studying the spectrum associated with
real-valued solutions. These will be denoted by $U_0(x)$, and they will be
realized as solutions of the ODE
\[
-\frac{1}{2}\partial_{xx}U_0+V(x)U_0+U_0^3-\mu U_0=0.
\]
The solutions will have the property that $U_0(x)\to0$ exponentially fast as
$|x|\to+\infty$.

The eigenvalue problem associated with the real-valued solution will be
exactly that as given in \eref{e:32}, except that now $\calL$ will be
diagonal, i.e., $\calL=\diag(\calL_+,\calL_-)$, with
\[
\calL_+=-\frac12\partial_x^2-\mu+V(x)+3U_0^2,\quad
\calL_-=-\frac12\partial_x^2-\mu+V(x)+U_0^2.
\]
The fact that $\calL$ is diagonal, i.e., the eigenvalue problem is in the
canonical form
\[
\calL_+u=-\lambda v,\quad\calL_-v=\lambda u,
\]
means that not only can more be said about the Hamiltonian-Krein index for
the original problem, but the index for the associated pencil will also
change. This latter amendment follows from the fact that the pencil can be
formed directly without first passing to the intermediate stage of
``doubling-up" the eigenvalue problem.

Because the operators $\calL_\pm$
\begin{enumerate}
\item are Sturmian (e.g., see \citep[Chapter~2.3]{kapitula:sad13})
\item contain the unbounded potential $x^2$
\item contain a potential which decays exponentially fast as
    $|x|\to+\infty$,
\end{enumerate}
\noindent the general properties of $\sigma(\calL_\pm)$ are well-understood:
\begin{enumerate}
\item $\sigma(\calL_\pm)\subset\R$ is composed solely of point eigenvalues
\item each eigenvalue is geometrically and algebraically simple.
\end{enumerate}
The operator $\calL_+$ will (generically) be invertible, while $\calL_-U_0=0$
implies that $\dim[\ker(\calL_-)]=1$. If we let
$\Pi:X\mapsto\Span\{U_0\}^\perp$ denote the orthogonal (spectral) projection,
and set
\[
\calR=\Pi\calL_+\Pi,\quad\calS^{-1}=\Pi\calL_-\Pi,
\]
then the search for nonzero eigenvalues for the original problem is
equivalent to finding the spectrum of the pencil
\[
(\calR-z\calS)u=0,\quad z=-\lambda^2\quad\Rightarrow\quad
\calR u=-\lambda v,\,\,\calS^{-1}v=\lambda u.
\]

\begin{figure}
\begin{center}
\begin{tabular}{cc}
\includegraphics[width=6cm]{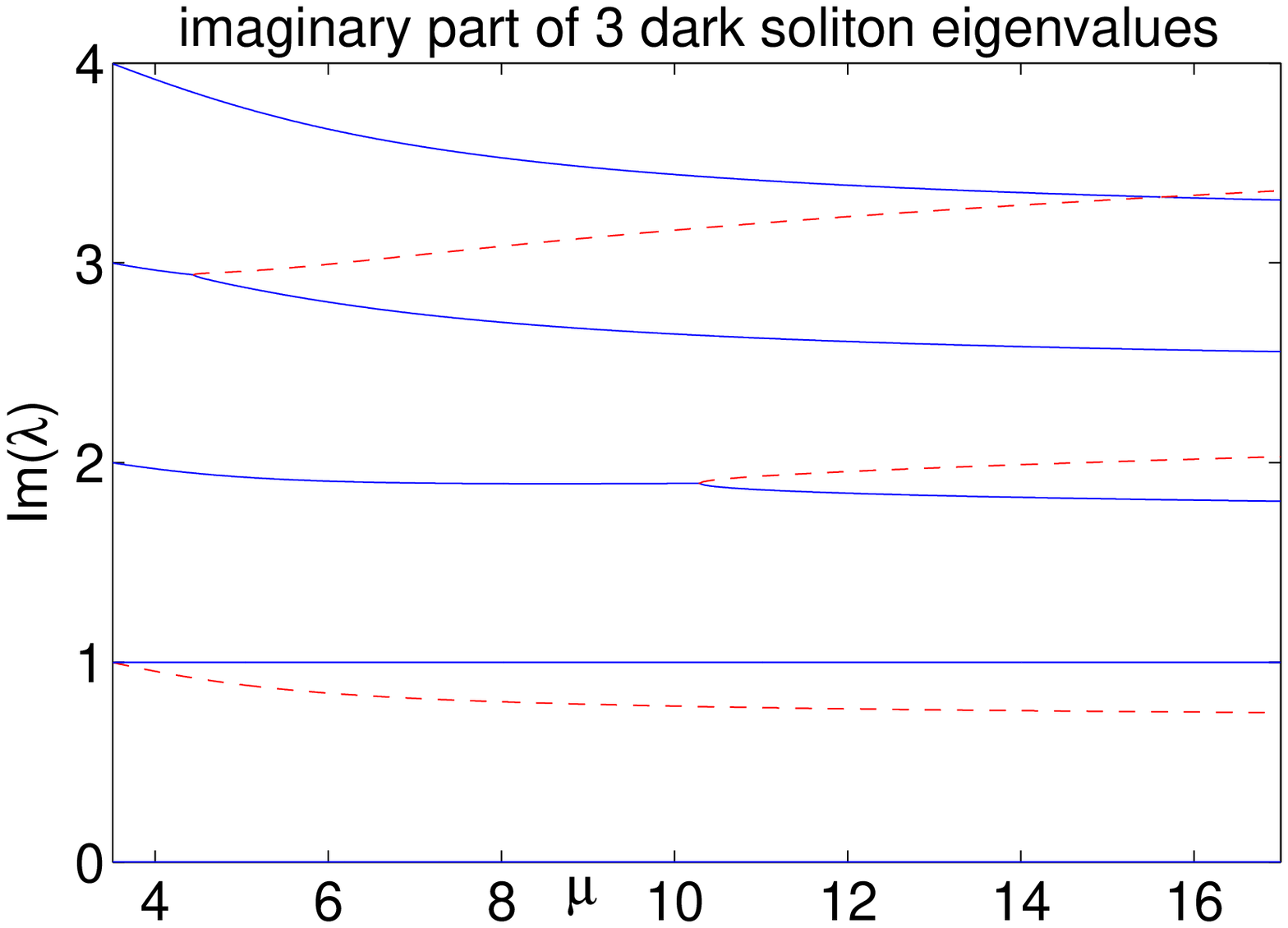} &
\includegraphics[width=6cm]{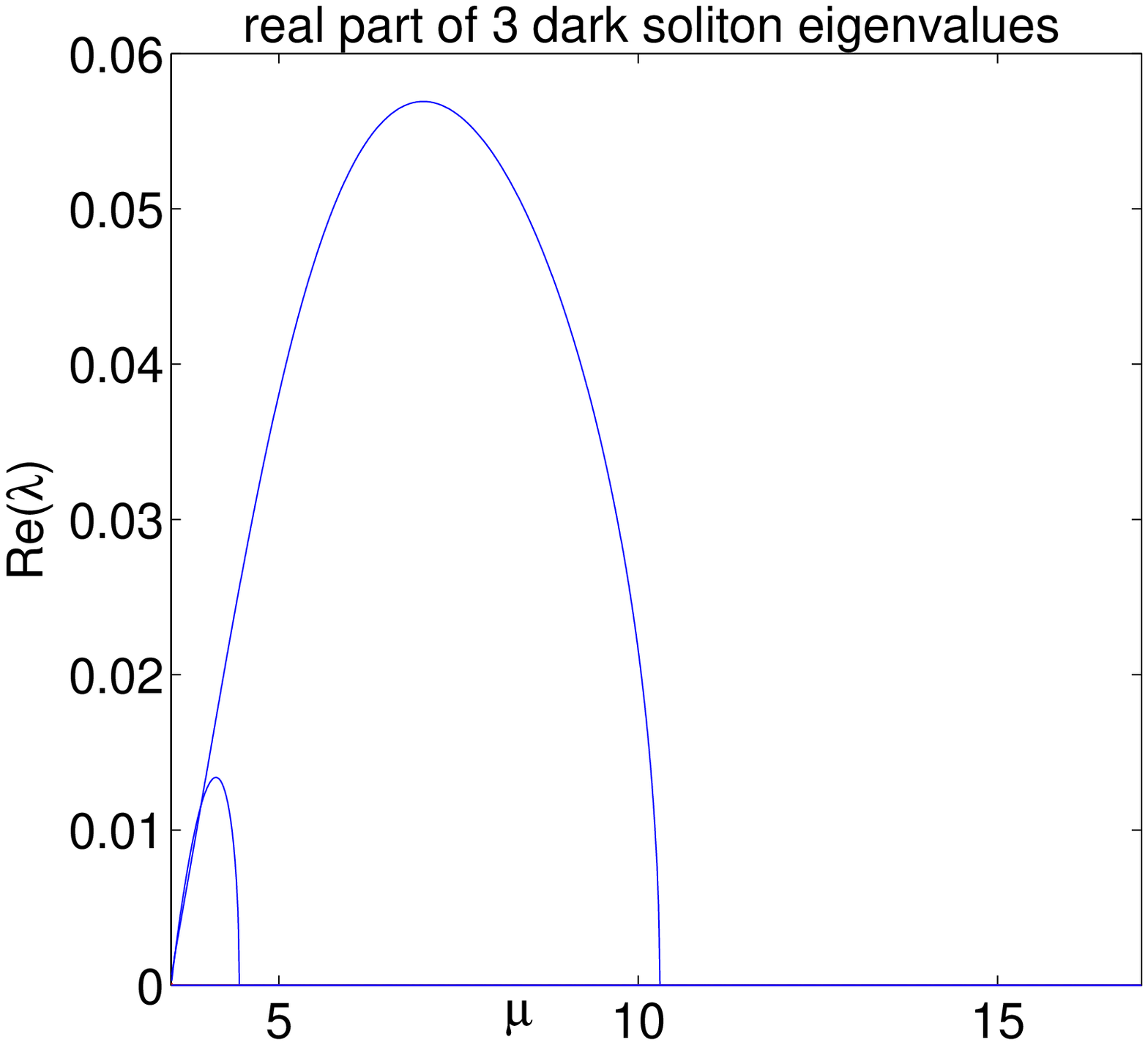} \\
\end{tabular}
\end{center}
\caption{(color online) Linear stability analysis for the 3-dark-soliton states: left panel is
$\Im(\lambda)$ versus $\mu$, while the right panel is the $\Re(\lambda)$ versus $\mu$.
The first critical value of $\mu$ where one pair of the complex eigenvalues becomes purely
imaginary is approximately $\mu=4.46$. The second critical value where the other
pair of complex eigenvalues becomes purely imaginary is approximately $\mu=10.29$. The eigenvalue(s)
which have negative Krein signature are shown as (red) dashed lines.}
\label{3_dark_eigen_lsa}
\end{figure}

The Hamiltonian-Krein index for the original eigenvalue problem is given by
\[
K_{\Ham}=\rmn(\calR)+\rmn(\calS)
\]
(compare with \eref{e:13}). Since $\Pi$ is a spectral projection, it will be
the case that $\rmn(\calS)=\rmn(\calL_-)$. As for $\rmn(\calR)$, by using the
Index Theorem in \citep{kapitula:tks10} it is the case that
\[
\rmn(\calR)=\rmn(\calL_+)-\rmn(\langle U_0,\calL_+^{-1}U_0\rangle).
\]
Upon using the fact that
\[
\calL_+(\partial_\mu U_0)=U_0\quad\Rightarrow\quad
\langle U_0,\calL_+^{-1}U_0\rangle=\frac12\partial_\mu\langle U_0,U_0\rangle
\]
we have the rewritten expression
\[
\rmn(\calR)=\rmn(\calL_+)-\rmn(\partial_\mu\langle U_0,U_0\rangle).
\]
In other words, the slope of the power curve $P(\mu)=\langle U_0,U_0\rangle$
has a direct influence on the number of negative directions of the operator
$\calL_+$ (this is the so-called Vakhitov–Kolokolov stability criterion for
canonical Hamiltonian eigenvalue problems). In conclusion, the
Hamiltonian-Krein index for the canonical problem is
\begin{equation}\label{e:42}
K_{\Ham}=\rmn(\calL_+)+\rmn(\calL_-)-\rmn(P'(\mu)),\quad
P(\mu)=\langle U_0,U_0\rangle.
\end{equation}

There is also an instability criterion. The classical result of
\citet{grillakis:lif88,jones:ios88} (recently reproven in
\cite{kapitula:sif12} using the Krein matrix) allows us to say that a lower
bound on $k_\rmr(\lambda)$ is caused by a difference in the number of
negative directions of $\calR$ and $\calS$. In other words,
\begin{equation}\label{e:43}
k_\rmr(\lambda)\ge|\rmn(\calL_-)-[\rmn(\calL_+)-\rmn(P'(\mu))]|,
\end{equation}
which in particular implies the previously stated result that
$k_\rmr(\lambda)\ge1$ if $|\rmn(\calL_-)-\rmn(\calL_+)|\ge2$.

\begin{figure}
\begin{center}
\includegraphics[width=9cm]{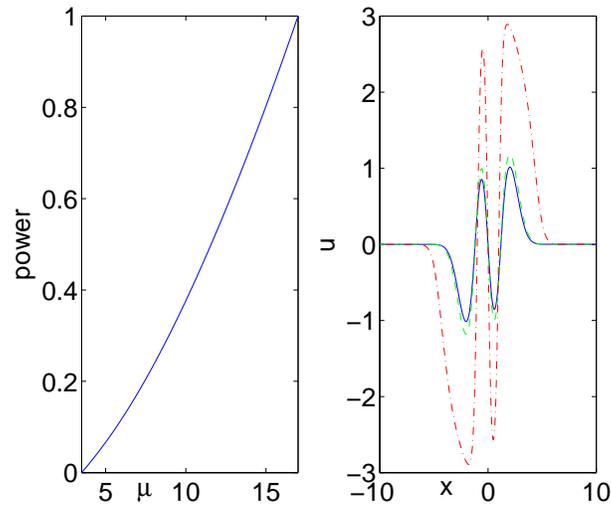}
\end{center}
\caption{(color online) The left panel shows the power $P(\mu)$ (see \eref{e:42}) versus the
chemical potential $\mu$: the power plotted here is scaled by the $\max(P(\mu))$. The right panel demonstrates the
stationary solutions as $\mu$ varies. The solid blue, dashed green, and dash-dotted red curves represent
the cases of $\mu=4.2, 4.6, 10.3$ respectively. Here the grid size is $n=800$ and the spatial interval is
$\Delta x=0.025$.}
\label{power_stationary_three_dark}
\end{figure}

Regarding the Hamiltonian-Krein index for the pencil, the fact that the
original system no longer needs to be ``doubled-up" in order to be put into
canonical form means that we just need to count eigenvalues with respect to
the eigenvalue mapping $z=-\lambda^2$ (see \autoref{f:SpectralPlaneMap}). In
particular,
\[
k_\rmr(\lambda)=k_\rmr(z),\quad k_\rmc(\lambda)=k_\rmc(z),\quad
k_\rmi^-(\lambda)=k_\rmi^-(z),
\]
which in the end means that the index for the pencil is precisely that for
the original problem, and is given in \eref{e:42}. Additionally, in the
construction of the Krein matrix the size no longer directly depends on
$K_{\Ham}$; instead, it will be of size $\rmn(\calS)=\rmn(\calL_-)$.

\begin{figure}
\begin{center}
\includegraphics[width=9cm]{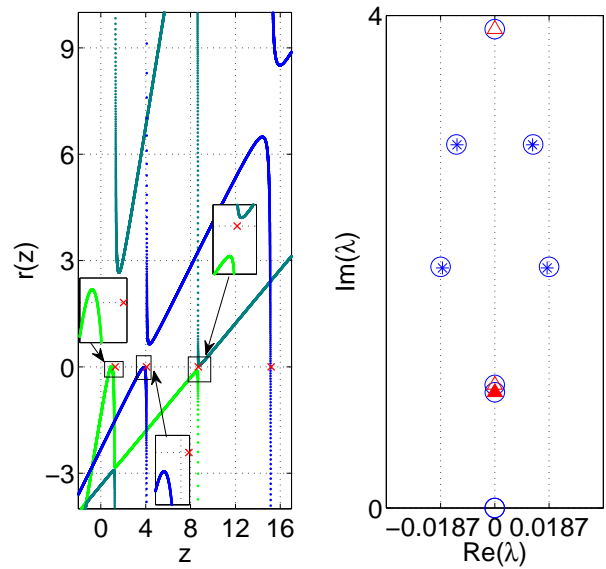}
\end{center}
\caption{(color online) The spectral picture when $\mu=4.2$, for which $k_\rmc(\lambda)=2$ and $k_\rmi^-(\lambda)=1$.
In the left panel is the plot of the three Krein eigenvalues. As predicted by the theory,
there is only one positive zero of a Krein eigenvalue for which the slope is
positive. In the right panel the eigenvalues for $\calJ\calL$ are denoted by (blue) circles.
The (red) triangles are the eigenvalues of $\calJ\calL$ which are realized as zeros of the Krein
eigenvalues. The (blue) stars represent the complex eigenvalues with nonzero real part as found
by the zeros of the Krein eigenvalues. The eigenvalue with negative Krein signature is marked with
a (red) filled triangle. Note that this eigenvalue is close to an eigenvalue with positive Krein
signature, which implies that the system is close to a Hamiltonian-Hopf bifurcation point.}
\label{krein_matrix_mu_42_3_dark}
\end{figure}

In the previous section we considered the spectral stability of solutions for
which two Krein eigenvalues could be used to locate the spectrum. We now
consider an example for which three Krein eigenvalues are needed. The
steady-state solution $U_0$ to be considered, hereafter called a
3-dark-soliton, is realized as a continuation of the Gauss-Hermite function
$(8x^3-12x)\rme^{-x^2/2}$ from the chemical potential $\mu^*=3.5$ (recall
that we assume $\Omega=1$). Let us first determine the Hamiltonian-Krein
index associated with this solution. A weakly nonlinear analysis along the
lines of that presented in, e.g., \citep{kapitula:tia06}, shows that for
$|\mu-\mu^*|$ small, the power satisfies the relationship $P'(\mu)>0$ (the
details are left for the interested reader). As we see from the numerics (see
\autoref{power_stationary_three_dark}), this relationship holds for all
values of $\mu$ under consideration. Regarding the indices of the operators
$\calL_\pm$, the combination of $U_0$ having three zeros and Sturm-Liouville
theory implies that $\rmn(\calL_-)=3$; consequently, in the Krein matrix
analysis there will be three Krein eigenvalues to follow. Regarding the
operator $\calL_+$, it is the case that in the weakly nonlinear limit
$\rmn(\calL_+)=3$: again, this relationship holds for all $\mu$ under
consideration. Appealing to \eref{e:42} we see that $K_{\Ham}=6$;
unfortunately, the lower bound of \eref{e:43} on $k_\rmr(\lambda)$ leads to
no definitive conclusion. Regarding the types of (potentially) unstable
eigenvalues, we have the following possibilities:
\begin{center}
\begin{tabular}{|c||c|c|c|c|c|c|c|c|c|c|}\hline
\rule[-3mm]{0mm}{8mm}$k_\rmr(\lambda)$   &0&0&0&0&2&2&2&4&4&6\\ \hline
\rule[-3mm]{0mm}{8mm}$k_\rmc(\lambda)$   &0&1&2&3&0&1&2&0&1&0\\ \hline
\rule[-3mm]{0mm}{8mm}$k_\rmi^-(\lambda)$ &3&2&1&0&2&1&0&1&0&0\\ \hline
\end{tabular}
\end{center}

\begin{figure}
\begin{center}
\includegraphics[width=9cm]{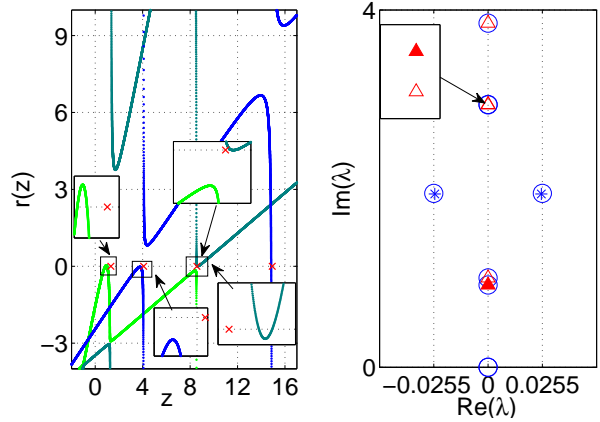}
\end{center}
\caption{(color online) Similar to \autoref{krein_matrix_mu_42_3_dark}, except now $\mu=4.46$ and
$k_\rmc(\lambda)=1$ with $k_\rmi^-(\lambda)=2$.}
\label{krein_matrix_mu_446_3_dark}
\end{figure}

\begin{figure}
\begin{center}
\includegraphics[width=9cm]{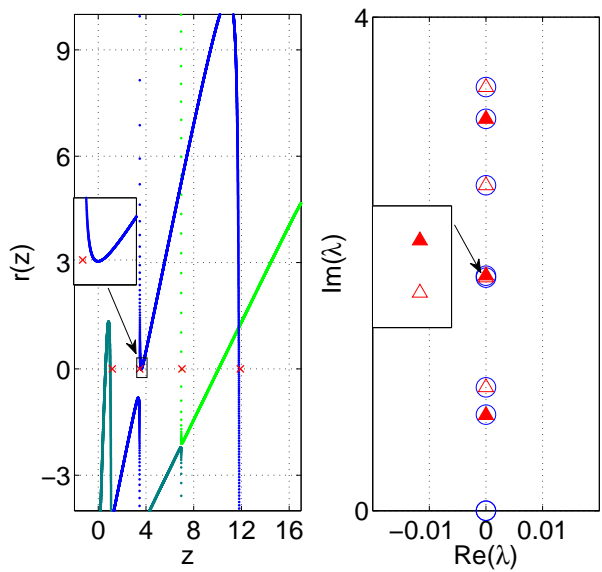}
\end{center}
\caption{(color online) Similar to \autoref{krein_matrix_mu_42_3_dark}, except now $\mu=10.30$
and $k_\rmi^-(\lambda)=3$.}
\label{krein_matrix_mu_1030_3_dark}
\end{figure}

In \autoref{3_dark_eigen_lsa} we show the location of the first few
eigenvalues from the linear stability analysis as a function of $\mu$. It is
always the case that $k_\rmr(\lambda)=0$, while $k_\rmc(\lambda)$ decreases
from two to one to zero. The decreasing of $k_\rmc(\lambda)$ implies that
$k_\rmi^-(\lambda)$ increases from one to two to three. The critical values
of $\mu$ for which $k_\rmc(\lambda)$ decreases are approximately
$\mu\sim4.465$ and $\mu\sim10.29$. In \autoref{krein_matrix_mu_42_3_dark} we
show the plot of the Krein eigenvalues (left panel) and full linearization
spectrum (right panel) when $\mu=4.2$, when $k_\rmc(\lambda)=2$ and
$k_\rmi^-(\lambda)=1$. The eigenvalue in the upper half-plane with negative
Krein signature is denoted with a (red) filled triangles. In
\autoref{krein_matrix_mu_446_3_dark} we show the plot of the Krein
eigenvalues (left panel) and full linearization spectrum (right panel) when
$\mu=4.46$, when $k_\rmc(\lambda)=1$ and $k_\rmi^-(\lambda)=2$. The two
eigenvalues in the upper half-plane with negative Krein signature are denoted
with (red) filled triangles. Finally, in \autoref{krein_matrix_mu_1030_3_dark} we show
the plot of the Krein eigenvalues (left panel) and full linearization
spectrum (right panel) when $\mu=10.30$, when $k_\rmi^-(\lambda)=3$. The
three eigenvalues in the upper half-plane with negative Krein signature are
denoted with (red) filled triangles.

\begin{remark}
In general, when considering $N$-dark-solitons to \eref{GPE_1d} it will be
the case that $P'(\mu)>0$ with $\rmn(\calL_+)=\rmn(\calL_-)=N$. Consequently,
when looking for the $2N$ (potentially) unstable eigenvalues for the
linearized problem it will be the case that $N$ Krein eigenvalues must be
plotted.
\end{remark}


\section{Conclusion}\label{s:5}

In the present work we have revisited Hamiltonian skew-adjoint eigenvalue
problems that typically arise in the linearization around a stationary state
of a Hamiltonian nonlinear partial differential equation. We presented a
brief overview of the known facts for the eigenvalue counts of the
corresponding (potentially) unstable spectra. We especially focused on a
novel, but straightforward, plan to implement finite dimensional techniques
to locate this spectrum via the singular points of the meromorphic Krein
matrix. We illustrate the value of the approach by considering realistic
problems for recently observed experimentally multi-vortex and multi-soliton
solutions in atomic Bose-Einstein condensates. We believe that this approach
can provide a valuable alternative to the highly-demanding computations that
require a diagonalization of the linearization matrix, especially in two- and
three-dimensional settings.

Naturally, there are numerous possibilities for further exploration. It would
be especially relevant for the atomic physics applications to use the present
method to examine the details of the spectra of three-dimensional solutions
such as vortex rings~\cite{komineas}. On the other hand, from a
methodological perspective, it would be also relevant to consider this
approach for the case of solutions of other classes of Hamiltonian problems
such as the Korteweg-de Vries equations and its generalizations, or nonlinear
Klein-Gordon equations. The latter also offer possibilities (e.g. in the
realm of stability of traveling waves etc.) to consider quadratic pencils
instead of linear ones, whereby extensions of the Krein matrix method would
be particularly desirable to develop from a rigorous mathematical viewpoint.
These themes are currently under study and will be reported in future
publications.




\phantomsection                                         
\addcontentsline{toc}{section}{\refname}                



\end{document}